\documentclass[aps,prd,preprint,groupedaddress,showpacs]{revtex4}

\usepackage{epsfig}
\usepackage{graphics}
\usepackage{amssymb,amsmath}
\usepackage{color}

\newcommand{\tr}{\mathop{\mbox{tr}}\nolimits}

\begin{document}

\leftmargin -2cm
\def\choosen{\atopwithdelims..}

~~\\
 DESY~14--050 \hfill ISSN 0418-9833
\\April 2014


 \boldmath
\title{Prompt-photon plus jet associated photoproduction at HERA\\ in the
parton Reggeization approach} \unboldmath

\author{\firstname{B.A. }\surname{Kniehl}}
\email{kniehl@desy.de}\affiliation{{II.} Institut f\"ur Theoretische Physik, Universit\" at Hamburg,
Luruper Chaussee 149, 22761 Hamburg, Germany}

\author{\firstname{M.A. }\surname{Nefedov}}
\email{nefedovma@gmail.com} \affiliation{Samara State University,
Academic Pavlov St.~1, 443011 Samara, Russia} \affiliation{{II.} Institut
f\"ur Theoretische Physik, Universit\" at Hamburg, Luruper Chaussee
149, 22761 Hamburg, Germany}

\author{\firstname{V.A.} \surname{Saleev}} \email{saleev@samsu.ru}
\affiliation{Samara State University, Academic Pavlov St.~1, 443011 Samara,
Russia} \affiliation{S.P.~Korolyov Samara State Aerospace
University, Moscow Highway 34, 443086 Samara, Russia}

\begin{abstract}
We study the photoproduction of isolated prompt photons associated with hadron
jets in the framework of the parton Reggeization approach.
The cross section distributions in the transverse energies and pseudorapidities
of the prompt photon and the jet as well as the azimuthal-decorrelation
variables measured by the H1 and ZEUS collaborations at DESY HERA are nicely
described by our predictions.
The main improvements with respect to previous studies in the
$k_T$-factorization framework include the application of the Reggeized-quark
formalism, the generation of exactly gauge-invariant amplitudes with off-shell
initial-state quarks, and the exact treatment of the $\gamma R\to \gamma g$ box
contribution with off-shell initial-state gluons.
\end{abstract}

\maketitle

\section{Introduction}
\label{sec:intro}

The photoproduction of prompt photons with large transverse momenta provides a
formidable laboratory for precision tests of perturbative quantum
chromodynamics (QCD) and a useful source of information on the parton content
of the proton and the real photon.
The initial-state photon may interact with the partons inside the proton either
directly (direct photoproduction) or via its partonic content
(resolved photoproduction).

The inclusive photoproduction of prompt photons, singly and in association with
jets, received a lot of attention, both experimentally and theoretically.
On the experimental side, the H1~\cite{H1_data1,H1_data2} and
ZEUS~\cite{ZEUS_data1,ZEUS_data2,ZEUS_data3} collaborations measured the
cross section distributions in the transverse energies $(E_T)$ and the
pseudorapidities $(\eta)$ of the prompt photon and the jet as well as in
azimuthal-decorrelation parameters such as the azimuthal angle enclosed between
the prompt-photon and jet transverse momenta ($\Delta\phi$) and the component
of the prompt-photon transverse momentum orthogonal to the direction of the jet
transverse momentum ($p_\perp$).
Also, the distributions in the variables estimating the momentum fractions of
the initial-state partons, $x^{\rm LO}_p$, $x^{\rm LO}_\gamma$, and
$x^{\rm obs}_\gamma$, were measured.
This rich set of observables allows one to perform a detailed study of the
underlying partonic processes and to assess the relevance of different
perturbative corrections.

On the theoretical side, attempts to describe this data were made both at
next-to-leading order (NLO) in the conventional collinear parton model (CPM)
\cite{NLO_FGH,NLO_KZ} and in approaches accommodating off-shell initial-state
partons, such as the $k_T$-factorization approach (KFA) \cite{LZ1,LZ2,MLZ} and
its implementation with Reggeized partons, which we refer to as the parton
Reggeization approach (PRA) \cite{Sal_prompt_photon_HERA}.
In the case of inclusive prompt-photon photoproduction, both the NLO CPM and
leading-order (LO) KFA predictions underestimate all the measured
distributions, as may be seen, e.g., from the comparative figures in
Refs.~\cite{H1_data1,H1_data2,ZEUS_data2}, while the LO PRA predictions
describe the $E_T$ distributions quite well and the $\eta$ distributions
reasonably well \cite{Sal_prompt_photon_HERA}.

As for prompt-photon plus jet associated photoproduction, NLO CPM predictions
generally agree with the measured $\eta$ distributions, slightly underestimate
the $E_T$ distributions, and provide a poor description of the
azimuthal-decorrelation observables \cite{H1_data1,H1_data2}, due to the fact
that these distributions collapse to delta functions at LO in the CPM and,
therefore, strongly depend on the radiation of additional partons.
The available KFA predictions provide a better description of the measured
$E_T$ distributions and azimuthal-decorrelation observables, but are
implemented with matrix elements that manifestly violate gauge invariance,
which renders the quantitative improvements of the predictions questionable.
Furthermore, in the early studies \cite{LZ1,LZ2}, the partonic subprocess
pertaining to the scattering of a photon and an off-shell gluon,
$\gamma g^*\to \gamma g$, was not taken into account.
Later, this contribution was found to be numerically significant \cite{MLZ},
due to the large gluon luminosity under HERA conditions.
But the treatment of this contribution was approximate because the virtuality
of the initial-state gluon was not taken into account at the amplitude level,
but only in the kinematics of the process \cite{MLZ}.

In view of the shortcomings of the previous calculations mentioned above, it is
an urgent matter to perform an improved analysis of prompt-photon plus jet
associated photoproduction in the PRA, which allows one to treat off-shell
initial-state quarks and gluons in a gauge-invariant way.
Moreover, it is crucial to include the full dependence on the transverse
momentum of the off-shell (Reggeized) initial-state gluon $R$ in the process
$\gamma R\to \gamma g$.
These are two main goals of the present paper.

This paper has the following structure.
In Sec.~\ref{sec:tree}, a basic introduction to the PRA, a list of the
relevant partonic subprocesses, and the amplitudes for the tree-level
contributions are presented.
In the Sec.~\ref{sec:box}, the calculation of the one-loop amplitude of the
partonic subprocess $\gamma R\to \gamma g$ is discussed in some detail, and the
cross-checks applied to the results obtained are described.
A compact expression of this amplitude is presented in Appendix~\ref{sec:appA}.
The results of the numerical calculations and comparisons with experimental
data and previous studies are carefully discussed in Sec.~\ref{sec:results},
and a few concluding remarks are collected in Sec.~\ref{sec:concl}.

\section{PRA formalism and tree-level contributions}
\label{sec:tree}

In hadron-hadron or lepton-hadron collisions with large center-of-mass
energies $\sqrt{S}$, different kinds of perturbative corrections are relevant
for different processes and different regions of phase space.
For example, the higher-order corrections for the production of heavy final
states, such as Higgs bosons, top-quark pairs, dijets with large invariant
masses, or Drell-Yan pairs, by initial-state partons with relatively large
momentum fractions $x\sim 10^{-1}$ are dominated by soft and collinear gluons
and may increase the cross sections up to a factor of two.

By contrast, relatively light final states, such as small-transverse-momentum
heavy quarkonia, single jets, prompt photons, or dijets with small invariant
masses, are produced by the fusion of partons with small values of $x$,
typically $x\sim 10^{-3}$, because of the large values of $\sqrt{S}$.
Radiative corrections to such processes are dominated by the production of
additional hard jets.
The only way to treat such processes in the conventional CPM is to calculate
higher-order corrections in the strong coupling constant $\alpha_s=g_s^2/4\pi$,
which could be a challenging task for some processes even at the NLO level.
To overcome this difficulty and take into account a sizable part of the
higher-order corrections in the small-$x$ regime, the KFA, also known as
high-energy factorization approach, was introduced \cite{kTf}.
The KFA works with off-shell initial-state partons, which carry not only a
fraction $x$ of longitudinal momentum, but also a significant transverse
momentum ${\bf q}_{T}$, with
$|{\bf q}_{T}|\sim x\sqrt{S}$.
The corresponding factorization formula may be schematically represented as
  \begin{equation}
  d\sigma({\cal Y})=\sum\limits_{i,j}\Phi_{i}(x_1,t_1,\mu_F)\otimes\Phi_{j}(x_2,t_2,\mu_F)\otimes d\hat{\sigma}_{i j}
  (x_1,{\bf q}_{T1}; x_2,{\bf q}_{T2},{\cal Y}), \label{kTf_sigma}
  \end{equation}
where the sum runs over the parton species $i$, $j$, $\otimes$ denotes a
convolution over the relevant momentum components of the partons, ${\cal Y}$ is
the set of kinematic variables of the final state, and $d\hat{\sigma}_{ij}$ are
the partonic cross sections.
The unintegrated parton distribution function (unPDF) $\Phi_{i}(x_i,t_i,\mu_F)$
depends on the longitudinal-momentum fraction $x_i$ and the virtuality
$t_i={\bf q}_{Ti}^2$ of the parton and the factorization scale $\mu_F$, which
separates the stages of the evolution of the unPDF and the hard scattering.
The unPDF is normalized by the following condition:
  \begin{equation}
  \int\limits^{\mu_F^2} dt\, \Phi_i(x,t,\mu_F)=xf_i(x,\mu_F), \label{Phi_NC}
  \end{equation}
where $f_i(x,\mu_F)$ is the respective CPM PDF.

In the asymptotic high-energy (Regge) regime, the characteristic scales of the
scattering process obey the following hierarchy:
$\Lambda_{\rm QCD}\ll\mu_F\sim \mu_R\ll \sqrt{S}$, where $\Lambda_{\rm QCD}$ is the
asymptotic scale parameter of QCD and $\mu_R$ is the renormalization scale.
Deeply in the Regge regime, all the produced particles are highly separated in
rapidity, obeying the so-called multi-Regge kinematics, while the $k_T$
ordering of the Dokshitzer-Gribov-Lipatov-Altarelli-Parisi (DGLAP) \cite{DGLAP}
evolution is completely broken.
So, the evolution of the unPDFs is governed by large logarithms of a new type,
namely $\log(1/x)$. 
To resum these logarithms, the Balitsky-Fadin-Kuraev-Lipatov (BFKL) evolution
equation was introduced \cite{BFKL}.
This leads to a powerlike growth of the gluon unPDF in the small-$x$ region,
while the effect on the quark unPDFs is subleading, so that the gluon unPDF is
expected to strongly dominate at high energies.

At intermediate energies, however, the quark unPDFs and DGLAP effects cannot
be neglected, and the unPDFs may be obtained by certain approaches taking into
account both DGLAP and BFKL effects, for example by the
Kimber-Martin-Ryskin approach \cite{KMR} or by a solution to the
Ciafaloni-Catani-Fiorani-Marchesini evolution equation \cite{CCFM}.


Special care is required to define hard-scattering matrix elements in the KFA
because initial-state partons are now off shell, which generally entails gauge
dependence in QCD.
In the KFA studies of heavy-quark pair production or deep-inelastic scattering,
the polarization vector of the initial-state gluon with four-momentum
$k^\mu=(k_0,{\bf k}_T,k_z)$ is usually taken to be
  \begin{equation}
  \varepsilon^\mu(k)=\frac{k_T^\mu}{|{\bf k}_T|},
\label{kTf_pr}
  \end{equation} 
where $k_T^\mu=(0,{\bf k}_T,0)$,
in analogy with the equivalent-photon approximation in QED \cite{kTf}.
However, this prescription does not lead to gauge-invariant results for
hard-scattering amplitudes with gluons in the final state because of their
involved non-abelian color structure.
Furthermore, the usual KFA does not provide a generally accepted prescription
for the treatment of off-shell initial-state quarks.
  
A rigorous way to solve this gauge-dependence problem is to observe that the
small-$x$ regime, with $x\sim \mu_F/\sqrt{S}\ll 1$, implies that particles
produced in the hard interaction are strongly separated in rapidity from the
particles produced at the unPDF evolution stage.
The regime where the produced particles are grouped in a few clusters which are
strongly separated in rapidity is characterized by what is called
quasi-multi-Regge kinematics (QMRK).
It was shown \cite{QMRK, QMRKrev} that, in the QMRK, the gauge-invariance
conditions hold for each cluster separately and that the fields carrying
four-momentum between these clusters are new gauge-invariant degrees of
freedom accompanying the usual Yang-Mills gluons and quarks in the effective
field theory for the Regge limit of QCD \cite{LipatovEFT}, the Reggeized
gluons \cite{LipatovEFT} and quarks \cite{LipVyaz}.
The implementation of the KFA, characterized by Eq.~(\ref{kTf_sigma}) with
partonic cross sections obtained using the Feynman rules of the effective field
theory for the Regge limit of QCD \cite{LipVyaz,FeynRules} is referred to as
the PRA.

The hard-scattering amplitudes in the PRA coincide with those obtained using
the prescription in Eq.~(\ref{kTf_pr}) whenever the application of the latter
is safe, as was explicitly shown, e.g., for
heavy-quark \cite{SVheavy} and heavy-quarkonium production
\cite{KSVcharm,KSVbottom}.
Recent examples of nontrivial applications of the PRA to high-energy
phenomenology include the description of dijet azimuthal decorrelations
\cite{dijets} as well as the production of bottom-flavored jets \cite{bjets},
Drell-Yan lepton pairs \cite{DY}, single jets, and prompt photons \cite{PPSJ}
at the Tevatron and the LHC.

We now turn from the general discussion of the relationship between the PRA and
the KFA to the application of the PRA to prompt-photon plus jet associated
photoproduction.
The LO QMRK approximation for this process corresponds to only including
$2\to 2$ subprocesses yielding potentially sizable contributions.
These partonic subprocesses may be classified into direct-photoproduction ones,
where the photon directly takes part in the hard scattering, and
resolved-photoproduction ones, in which the photon interacts as a composite
object containing quarks and gluons.
The LO direct-photoproduction subprocesses are
  \begin{eqnarray}
  Q(q_1)+\gamma(q_2) &\to & q(q_3)+\gamma (q_4), \label{dir_compton} \\
  R(q_1)+\gamma(q_2) &\to & g(q_3)+\gamma (q_4), \label{dir_box}
  \end{eqnarray}
where $Q$ and $R$ are the Reggeized quark and gluon from the proton and the
four-momenta of the partons are given in parentheses.
Here, the charge-conjugated subprocesses, involving the Reggeized antiquark
$\bar{Q}$, are also implied.
The contribution to the cross section of the partonic subprocess in
Eq.~(\ref{dir_compton}) is of order ${\cal O}(\alpha^2)$, where $\alpha$ is
Sommerfeld's fine-structure constant.
The contribution from the partonic subprocess in Eq.~(\ref{dir_box}) is
formally of order ${\cal O}(\alpha^2\alpha_s^2)$.
However, due to the large values of the gluon unPDF at small values of $x$,
this process should be taken into account already at LO in the PRA.
The LO resolved-photoproduction subprocesses are
  \begin{eqnarray}
  R(q_1)+q\left[\gamma\right](\tilde{q}_2) &\to & q(q_3)+\gamma(q_4), \label{res_compton1} \\
  Q(q_1)+\bar{q}\left[\gamma\right](\tilde{q}_2) &\to & g(q_3)+\gamma (q_4), \label{res_annig1} \\
  Q(q_1)+g\left[\gamma\right](\tilde{q}_2) &\to & q(q_3)+ \gamma(q_4), \label{res_compton2}
  \end{eqnarray}
and their charge-conjugated counterparts.
The partonic subprocess in Eq.~(\ref{res_compton1}) is important because of the
above-mentioned amplification by the gluon unPDF of the proton.
By detailed inspection, we find the partonic subprocesses in
Eqs.~(\ref{res_annig1}) and (\ref{res_compton2}) to account for less than
5~\% of the total cross section and omit their contributions in the following.

In addition, partonic subprocesses in which final-state partons fragment to
photons should be considered.
However, their contribution is strongly suppressed by the photon-isolation
condition applied to the experimental data, which
constrains the hadronic energy within the photon isolation cone to be less than
$10\%$ of the photon energy.
In other words, more than $90\%$ of the parton energy must be transmitted to
the photon, which rarely happens.
We explicitly verify the strong supression of the fragmentation contributions,
which was also observed in Ref.~\cite{Sal_prompt_photon_HERA}.

All the $2\to 3$ subprocesses contribute at NLO in the PRA.
In order to avoid double counting of contributions to unPDFs and
hard-scattering matrix elements due to the emission of additional partons, one
may impose the condition that there are no rapidity gaps between unobserved and
observed partons, which requires a proper subtraction procedure, as described
in Ref.~\cite{substr}.
Contributions of this type and from the interference of one-loop and
tree-level $2\to 2$ scattering amplitudes constitute non-factorizable
higher-order corrections in our approach, in contrast to those which can be
factorized into unPDFs.
In the present paper, we focus on the LO contributions.

In the remainder of this section, we outline the derivation of the amplitudes
for the tree-level subprocesses in Eqs.~(\ref{dir_compton}) and
(\ref{res_compton1}).
We start by introducing the basic kinematic notation to be used throughout
this paper.
We work in the laboratory frame and take the $z$ axis to point along the flight
direction of the proton, whose mass we neglect.
It is convenient to introduce the light-cone four-vectors
\begin{equation}
 n^\mu_+=\frac{P_2^\mu}{E_2},\qquad n_-^\mu=\frac{P_1^\mu}{E_1},
\end{equation}
where $P_1$ and $P_2$ are the four-momenta of the proton and the electron,
respectively, and $E_1$ and $E_2$ are their energies.
We have $n_\pm^2=0$ and $n_+\cdot n_-=2$.
Then, any four-vector $k^\mu$ may be expressed in terms of its light-cone
components, $k^\pm=n_\pm\cdot k=k^0\pm k^3$, as
\begin{eqnarray}
k^\mu=\frac{1}{2}\left(k^+n_-^\mu+k^-n_+^\mu\right)+k_T^\mu,
\end{eqnarray}
and we have $k_T\cdot n_\pm=0$.
The four-momentum of the Reggeized parton from the proton can be written as
$q_1=x_1P_1+q_{T1}$ and has virtuality $q_1^2=q_{T1}^2=-{\bf q}_{T1}^2=-t_1$.
The quasi-real photon carries the fraction $y$ of the electron energy and has
four-momentum $q_2=yP_2$.
If the photon is resolved, then it transfers the fraction $x_2$ of its energy
to the offspring parton, which has four-momentum $\tilde{q}_2=x_2q_2=x_2yP_2$.
In the following, we assume the photon to be direct; the resolved-photon
results are recovered by replacing $q_2$ with $\tilde{q}_2$.
The square of the proton-photon center-of-mass energy is
$S=2P_1\cdot q_2=4yE_1E_2$.
The partonic Mandelstam variables are defined as
\begin{equation}
s=(q_3+q_4)^2,\qquad
t=(q_2-q_4)^2,\qquad
u=(q_2-q_3)^2,
\label{mandel}
\end{equation}
where $q_3$ and $q_4$ are the four-momenta of the final-state particles, which
we take to be massless, and we have $s+t+u=-t_1$.
They may be expressed in terms of the final-state light-cone four-momenta as
  \begin{equation}
  s=(q_3^+ + q_4^+)(q_3^- + q_4^-)-t_1,\qquad
  t=-q_4^+(q_3^-+q_4^-),\qquad
  u=-q_3^+(q_3^-+q_4^-). \label{LC-Man}
  \end{equation}
It turns out that the hard-scattering amplitudes may be cast into a
particularly compact form by using the dimensionless Sudakov variables instead
of the light-cone ones.
They are defined as
\begin{equation}
  a_{3,4}=\frac{2q_2\cdot q_{3,4}}{S}=\frac{2yE_2 q_{3,4}^+}{S},\qquad
  b_{3,4}=\frac{2P_1\cdot q_{3,4}}{S}=\frac{2E_1 q_{3,4}^-}{S},
\end{equation}
so that $a_3+a_4=x_1$ and $b_3+b_4=1$, or $b_3+b_4=x_2$ in the resolved-photon
case.

\begin{figure}[H]
\includegraphics[width=0.5\textwidth]{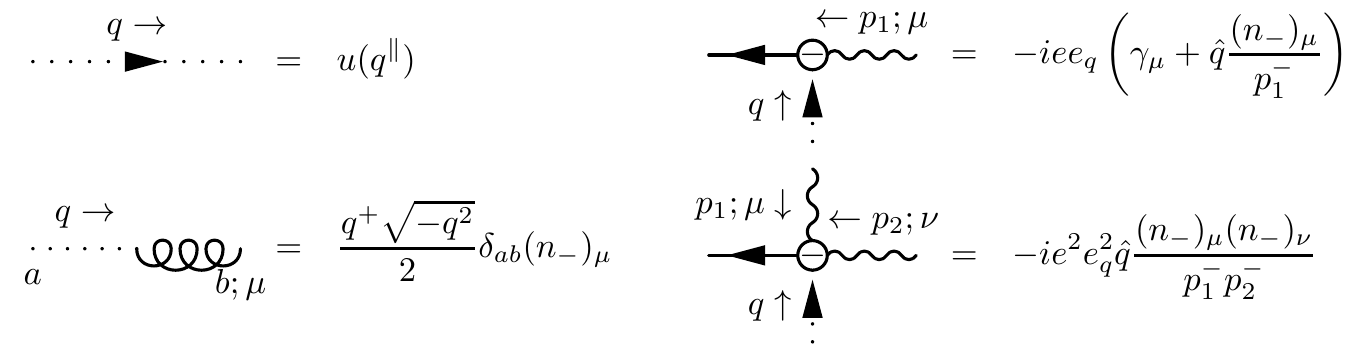}
\caption{\label{fig_FRs}
Feynman rules for the Reggeized quark and gluon in the initial state (left
panel) and for the $Q\gamma q$ and $Q\gamma\gamma q$ vertices (right panel).}
\end{figure}

In addition to the standard Feynman rules of QCD, we need the couplings of the
Reggeized quarks and gluons to the ordinary quarks, gluons, and photons.
The full list of the latter may be found in Refs.~\cite{LipVyaz,FeynRules}.
For the reader's convinience, we specify the Feynman rules relevant for our
calculation in Fig.~\ref{fig_FRs}.

\begin{figure}[H]
\includegraphics[width=0.5\textwidth]{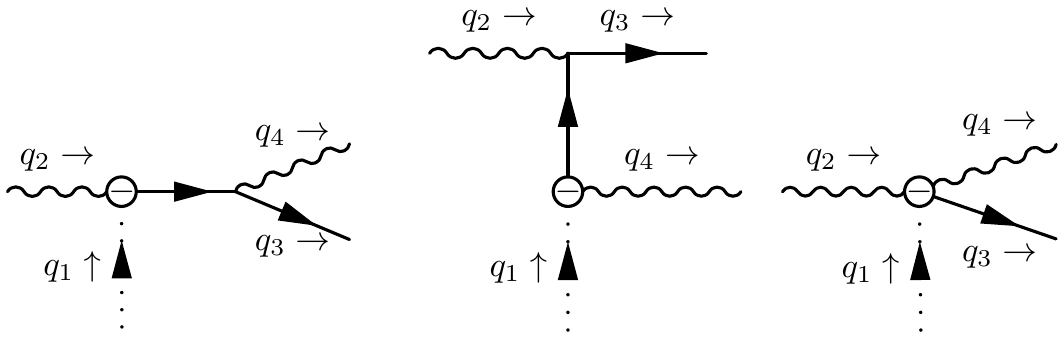}
\caption{\label{fig_Dir_compton_Ds}%
Feynman diagrams contributing at LO to the partonic subprocess in
Eq.~(\ref{dir_compton}).}
\end{figure}

The Feynman diagrams contributing at LO to the partonic subprocess in
Eq.~(\ref{dir_compton}) are shown in Fig.~\ref{fig_Dir_compton_Ds}.
Using the Feynman rules in Fig.~\ref{fig_FRs} and the light-cone four-vectors
defined above, we find the modulus square of the hard-scattering amplitude,
averaged over the spins and colors of the incoming partons and summed over
those of the outgoing ones, to be
  \begin{equation}
 \overline{\left\vert{\cal M}(Q\gamma\to q\gamma) \right\vert^2}
=-32\pi^2\alpha^2 e_q^4\frac{Sx_1}{b_4su}\left(t_1b_3^3+sb_4^3-u\right),
\label{A2_dir_compton}
  \end{equation}
where $e_q$ is the quark electric charge in the units of the positron charge.
In the limit when the initial-state Reggeized quark goes on shell, which
amounts to substituting $t_1\to 0$, $a_3\to-u/S$, $a_4\to -t/S$,
$b_3\to -t/(x_1S)$, $b_4\to -u/(x_1S)$, and $x_1\to s/S$,
Eq.~(\ref{A2_dir_compton}) reproduces the well-known LO CPM result for Compton
scattering,
\begin{equation}
\overline{\left\vert{\cal M}(q\gamma\to q\gamma) \right\vert^2}
=-32\pi^2\alpha^2e_q^4\left(\frac{s}{u}+\frac{u}{s}\right).
\end{equation}

\begin{figure}[H]
\includegraphics[width=0.5\textwidth]{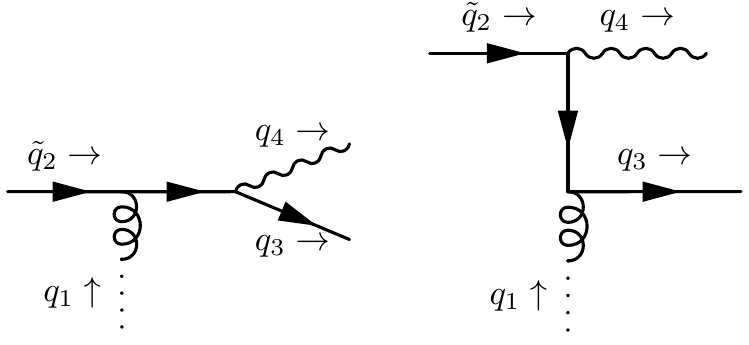}
\caption{\label{fig_Res_compton_Ds}%
Feynman diagrams contributing at LO to the partonic subprocess in
Eq.~~(\ref{res_compton1}).}
\end{figure}

The Feynman diagrams contributing to the partonic subprocess in
Eq.~(\ref{res_compton1}) are depicted in the Fig.~\ref{fig_Res_compton_Ds}.
The modulus square of the hard-scattering amplitude averaged over the spins
and colors in the initial state and summed over those in the final state reads
  \begin{eqnarray}
  \overline{\left\vert{\cal M}(Rq\to q\gamma) \right\vert^2}
&=&\frac{16}{3}\pi^2 \alpha\alpha_s e_q^2\frac{S^2 x_1^2 x_2}
  {st^2t_1}\left\lbrace t\left[ub_3+(t+u)b_4-Sa_3b_3^2+sx_2\right]\right.
\nonumber\\
 &&{}+ \left. Sa_4b_3\left[sb_4-tb_3-(s+t)x_2\right] \right\rbrace.
\label{A2_res_compton}
  \end{eqnarray}
The CPM limit of Eq.~(\ref{A2_res_compton}) is defined as
  \begin{equation}
  \lim\limits_{t_1\to 0} \int\limits_0^{2\pi}\frac{d\phi_1}{2\pi}
 \overline{\left\vert{\cal M}(Rq\to q\gamma)
   \right\vert^2} 
= \overline{\left\vert{\cal M}(gq\to q\gamma) \right\vert^2},
\label{CL_def}
  \end{equation}
where $\phi_1$ is the azimuthal angle enclosed between the three-vectors
${\bf q}_{T1}$ and ${\bf q}_{T3}$.
Note that the order of integrating and taking the limit may be safely reversed
in Eq.~(\ref{CL_def}).
The limit $t_1\to 0$ may be taken in Eq.~(\ref{A2_res_compton}) by
substituting $a_3\to -u/(Sx_2)$, $a_4\to -t/(Sx_2)$,
$b_3\to (-t+B\sqrt{t_1})/(Sx_1)$, $b_4\to (-u -B\sqrt{t_1})/(Sx_1)$, and
$Sx_1x_2\to s+t_1$, where $B=\sqrt{2ut/s}\cos\phi_1$.
We thus recover the well-known LO CPM result
\begin{equation}
\overline{\left\vert{\cal M}(gq\to q\gamma) \right\vert^2}=-\frac{16}{3}\pi^2\alpha\alpha_se_q^2\left(\frac{s}{t}+\frac{t}{s}\right).
\end{equation}

Recently, an alternative method, which is equivalent to the PRA involving
Reggeized quarks and gluons adopted here, was proposed in Ref.~\cite{HKS}.
It amounts to embedding the $2\to n$ scattering processes under consideration
here into auxiliary $2\to n+2$ scattering processes and to extracting from them
the gauge-invariant $2\to n$ amplitudes with off-shell initial-state partons by
using the spinor-helicity representation with complex momenta.
This is more suitable for the implementation in automatic matrix-element
generators, but the use of Reggeized quarks and gluons is by far simpler for
the purposes of the present study.

\section{Box contribution}
\label{sec:box}

In this section, we discuss the hard-scattering amplitude of the one-loop
subprocess in Eq.~(\ref{dir_box}) within the PRA.
Specifically, we derive the helicity amplitudes and verify that they reproduce
the well-known expressions for photon-by-photon scattering~\cite{BoxCPM} in the
CPM limit.

\begin{figure}[H]
\includegraphics[width=0.5\textwidth]{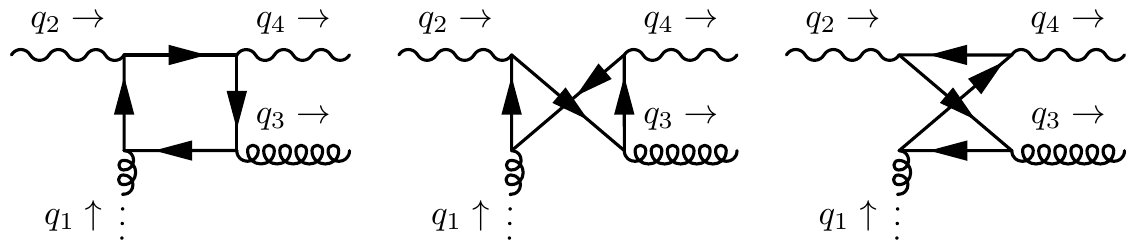}
\caption{\label{fig_Dir_Box_Ds}%
Feynman diagrams contributing at LO to the partonic subprocess in
Eq.~(\ref{dir_box}).
The diagrams with reversed fermion-number flow are not shown.}
\end{figure}

The contributing Feynman diagrams are shown in Fig.~\ref{fig_Dir_Box_Ds}.
Using the Feynman rules in Fig.~\ref{fig_FRs}, we may write the helicity
amplitudes as
   \begin{equation}
   {\cal M}(R \lambda_2, \lambda_3 \lambda_4)=-\frac{q_1^+}{2\sqrt{t_1}}(n_-)_{\mu_1}\varepsilon_{\mu_2}(1,-\lambda_2)
   \varepsilon^*_{\mu_3}(2,\lambda_3)\varepsilon^*_{\mu_4}(2,-\lambda_4) {\cal M}^{\mu_1\mu_2\mu_3\mu_4},\label{Rgamma}
   \end{equation}
where $\lambda_i=\pm 1$ are the helicities of the massless vector bosons and
the overall factor
\begin{equation}
   \frac{(4\pi)^2\alpha\alpha_s}{(2\pi)^4}\frac{\delta_{ab}}{2}\left(\sum\limits_q e_q^2\right),
\end{equation}
has been omitted on the right-hand-side of Eq.~(\ref{Rgamma}) for the ease of
notation.
The fourth-rank vacuum polarization tensor ${\cal M}^{\mu_1\mu_2\mu_3\mu_4}$
in Eq.~(\ref{Rgamma}) reads
   \begin{eqnarray}
   {\cal M}^{\mu_1\mu_2\mu_3\mu_4}&=&2\int d^4q\left\{
\frac{\tr\left[(\hat{q}-\hat{q}_1)\gamma^{\mu_3}(\hat{q}+\hat{q}_2-\hat{q}_4)
   \gamma^{\mu_4}(\hat{q}+\hat{q}_2)\gamma^{\mu_2}\hat{q}\gamma^{\mu_1}\right]}
{ (q-q_1)^2 (q+q_2-q_4)^2 (q+q_2)^2 q^2}\right.
\nonumber\\
&&{}+\left. (q_3\leftrightarrow q_4, \mu_3 \leftrightarrow \mu_4) +  (q_4\leftrightarrow -q_2, \mu_4 \leftrightarrow \mu_2)\right\},\label{m1m2m3m4}
   \end{eqnarray}
where the overall factor of two accounts for the Feynman diagram with the
fermion-number flow reversed.
The polarization four-vectors in Eq.~(\ref{Rgamma}) read
 \begin{equation}
   \varepsilon(j,\lambda)=\frac{1}{\sqrt{2}}\left(n^{(j)}_x
+i\lambda n^{(j)}_y \right),
\end{equation}
where
\begin{eqnarray}
   n^{(1)}_x&=& \frac{1}{\Delta}[(q_2\cdot q_3)q-(q\cdot q_3)q_2-(q\cdot q_2)q_3],\nonumber \\
   n^{(2)}_x&=& \frac{1}{\Delta}[(q_3\cdot q_4)q-(q\cdot q_4)q_3-(q\cdot q_3)q_4],\nonumber \\
   (n_y^{(1)})^\mu&=&-(n_y^{(2)})^\mu=\frac{1}{\Delta}\epsilon^{\mu q_2 q_3 q_4}
\equiv n_y^\mu,\nonumber
 \end{eqnarray}
with $\Delta=\sqrt{stu}/2$ and $q=q_2+q_3$.
The handling of the four-vector $n_y$ may be facilitated by observing that it
has the scalar products
  \begin{equation}
   q_2\cdot n_y=q_3\cdot n_y=q_4\cdot n_y=n_+\cdot n_y=0,\qquad n_y^2=-1,
   \end{equation}
and that the four-vector $n_-$ appearing in Eq.~(\ref{Rgamma}) may be
decomposed as
   \begin{equation}
   n_-=\alpha n_++\beta_1 q_3+\beta_2 q_4+\gamma n_y, \label{nmexp}\\
   \end{equation}
with the coefficients
  \begin{equation}
  \alpha=\frac{s_+-s}{2q_3^+q_4^+},\qquad
  \beta_1=\frac{s+s_-}{s q_3^+},\qquad
  \beta_2=\frac{s-s_-}{sq_4^+},\qquad
  \gamma=\frac{2yE_2}{\Delta}|{\bf q}_{T3}||{\bf q}_{T4}|\sin(\Delta\phi),
  \end{equation}
where $s_\pm=q_4^-q_3^+\pm q_4^+ q_3^-$ and $\Delta\phi$ is the azimuthal angle
enclosed between ${\bf q}_{T3}$ and ${\bf q}_{T4}$.
Exploiting the fact that $n_-^2=0$, we may express $\gamma^2$ through
light-cone components as
  \begin{equation}
  \gamma^2=\frac{2ss_+-s^2-s_-^2}{sq_3^+q_4^+} \label{gdef}.
  \end{equation}
In the CPM limit, the four-vectors $n_+$, $n_-$, $q_3$, and $q_4$ become
linearly dependent, and $\gamma\to0$.
For the sake of a compact expression for Eq.~(\ref{Rgamma}), we introduce the
variable
  \begin{equation}
  \gamma_1=\frac{q_3^+\Delta}{\sqrt{t_1}}\gamma =\frac{u}{\sqrt{t_1}}|{\bf q}_{T3}||{\bf q}_{T4}|\sin(\Delta\phi), \label{g1def}
  \end{equation}
which has a non-vanishing CPM limit,
  \begin{eqnarray}
  \gamma_1\to 2\frac{u}{s}\Delta\sin\phi_1. \label{ga1CL}
  \end{eqnarray}
A similiar variable, $\gamma_2$, is related to the product
\begin{equation}
q_3^-q_3^+=\frac{u}{(t+u)^2}\left[ (u-t)(t_1-t)-2t^2+\gamma_2\sqrt{t_1}\right].
\label{q3q3}
\end{equation}
It may be expressed through $\gamma_1$ using Eqs.~(\ref{LC-Man}), (\ref{gdef}),
and (\ref{g1def}) as
\begin{equation}
\gamma_2=2\zeta\sqrt{stu-\frac{(t+u)^2}{u^2}\gamma_1^2}, \label{g2def}
  \end{equation}
where the sign factor $\zeta=\pm 1$ is to be determined so that the product of
Eq.~(\ref{q3q3}) with $t$ always coincides with $uq_3^-q_4^+$ in compliance
with Eq.~(\ref{LC-Man}).
In the CPM limit, we have
 \begin{equation}
  \gamma_2\to 4\Delta\cos\phi_1.\label{ga2CL}
  \end{equation}
These new variables allow us to express Eq.~(\ref{Rgamma}) in a simple form
that is manifestly finite in the CPM limit.
All the dependences on the light-cone components resides in $\gamma_1$, while
the residual parts of the expression depend only on the Mandelstam variables.
The exact analytical expressions for all helicity amplitudes and the squared
amplitude are presented in Appendix~\ref{sec:appA} in terms of the
dimensionally-regularized one-loop scalar integrals $B_0$, $C_0$, and $D_0$
defined as in Ref.~\cite{K_Ellis}.
The cancellations of the ultraviolet and infrared divergences are explicit in
these expression and are also checked in the numerical calculations.

Another important consistency check is to recover the well-known result in the
CPM limit \cite{BoxCPM}.
The relationship analogous to Eq.~(\ref{CL_def}) may be written as
\begin{equation}
\int\limits_0^{2\pi}\frac{d\phi_1}{2\pi}\lim\limits_{t_1\to 0}
\overline{\left\vert{\cal M}(R\lambda_2,\lambda_3\lambda_4)\right\vert^2}
=\frac{|N|^2}{2}\sum_{\lambda_1=\pm}\left\vert{\cal M}
(\lambda_1\lambda_2,\lambda_3\lambda_4)\right\vert^2,
\label{CLbox}
\end{equation}
where the normalization factor $N=8\pi^2i$ has been pulled out of the CPM
amplitudes, so that
\begin{equation}
{\cal M}(++,+-)={\cal M}(++,-+)={\cal M}(+-,++)={\cal M}(-+,++)
={\cal M}(++,--)=-1.
\end{equation}
In the following, we set $\lambda_2=+1$ without loss of generality.
For $\lambda_3=\lambda_4=-1$, Eq.~(\ref{CLbox}) may be immediately verified
using Eqs.~(\ref{ga1CL}) and (\ref{HA4}).
The other three combinations of $\lambda_3$ and $\lambda_4$ are slightly more
involved.

Let us consider the case $\lambda_3=\lambda_4=+1$ as an example.
We first recall that \cite{BoxCPM}
\begin{equation}
{\cal M}(++,++)=1+\frac{u-t}{s}\left[B_0(t)-B_0(u)\right]+\frac{t^2+u^2}{s^2}\left[tC_0(t)+uC_0(u)-\frac{tu}{2}D_0(t,u)\right],
\end{equation}
where the short-hand notation for the scalar one-loop integrals $B_0$, $C_0$,
and $D_0$ is explained in Appendix~\ref{sec:appA}.
On the other hand, substituting Eqs.~(\ref{ga1CL}) and (\ref{ga2CL}) in
Eq.~(\ref{HA1}) and using $s+t+u=0$, we find
\begin{equation}
 \lim\limits_{t_1\to 0}{\cal M}(R+,++)=4\sqrt{2}\pi^2i
\left\{[1-{\cal M}(++,++)]e^{-i\phi_1}-2\cos\phi_1\right\}.
\label{yyy}
 \end{equation}
Taking the modulus squared of Eq.~(\ref{yyy}) and averaging over $\phi_1$, we
recover Eq.~(\ref{CLbox}) with $\lambda_2=\lambda_3=\lambda_4=+1$.
The residual two cases $\lambda_3=-\lambda_4=\pm1$ may be treated similarly.

We also check the CPM limit at the stage of numerical calculations by
temporarily adopting the following simple ansatz for the gluon unPDF:
 \begin{equation}
 \Phi_g(x,t,\mu_F)=xf_g(x,\mu_F)\frac{2}{\mu_F^2\sigma\sqrt{\pi}}
\exp\left(-\frac{t^2}{\mu_F^4\sigma^2}\right),
 \end{equation}
with a sufficiently small value of $\sigma$.
In fact, $\sigma\to 0$ corresponds to the CPM limit, in which the normalization
condition of Eq.~(\ref{Phi_NC}) is satisfied.

Finally, we recover our result for Eq.~(\ref{Rgamma}), including its full $t_1$
dependence, from the vector parts of the helicity amplitudes of the partonic
subprocess $gg\to Zg$ presented in Ref.~\cite{GloverGZ}.
To this end, we represent the projector $(n_-)_{\mu_1}$ in Eq.~(\ref{Rgamma})
as a linear combination of the transverse and longitudinal polarization
four-vectors of the $Z$ boson and perform a boost to the center-of-mass frame
used in Ref.~\cite{GloverGZ}.

\section{Numerical analysis}
\label{sec:results}

We are now in a position to present our numerical results for the cross section
of prompt-photon plus jet associated photoproduction in the PRA and to compare
them with HERA~II data \cite{H1_data1,H1_data2,ZEUS_data2,ZEUS_data3}.
We work in the laboratory frame, where the proton and electron have energies
$E_p=920$~GeV and $E_e=27.6$~GeV, respectively, and count rapidity positive in
the proton flight direction.
We call the transverse energies of the prompt photon and jet $E_T^\gamma$ and
$E_T^{\rm jet}$, their pseudorapidities $\eta^\gamma$ and $\eta^{\rm jet}$,
and their azimuthal angles $\phi^\gamma$ and $\phi^{\rm jet}$, respectively.
For the reader's convenience, we list our master formula for the hadronic cross
section differential in $E_T^\gamma$, $\eta^\gamma$, $E_T^{\rm jet}$,
$\eta^{\rm jet}$, $\Delta\phi=\phi^{\rm jet}-\phi^\gamma$, and $y$ defined
above Eq.~(\ref{mandel}),
\begin{equation}
\frac{d\sigma(pe\to\gamma+j+X)}{dE_T^\gamma d\eta^\gamma dE_T^{\rm jet}
d\eta^{\rm jet}d(\Delta\phi)dy}=\sum\limits_{i,j=q,\bar{q},g}
\Phi_i(x_1,t_1,\mu_F)G_{\gamma/e}(y)x_2f_{j/\gamma}(x_2,\mu_F)
\frac{E_T^\gamma E_T^{\rm jet}}{8\pi^2(yx_1x_2S_{pe})^2}
\overline{\left\vert {\cal M}_{ij}\right\vert^2},\label{sigma_res}
\end{equation}
where $S_{pe}=4E_pE_e$,
\begin{equation}
x_1=\frac{E_T^\gamma e^{\eta^\gamma}+E_T^{\rm jet}e^{\eta^{\rm jet}}}{2E_p},
\quad
x_2=\frac{E_T^\gamma e^{-\eta^\gamma}+E_T^{\rm jet}e^{-\eta^{\rm jet}}}
{2yE_e},
\quad
t_1=(E_T^\gamma)^2+(E_T^{\rm jet})^2+2E_T^\gamma E_T^{\rm jet}\cos(\Delta\phi).
\label{x1x2t1}
\end{equation}
In the Weizs\"{a}cker-Williams approximation \cite{WWappr}, the flux of
quasi-real photons is
\begin{equation}
G_{\gamma /e}(y)=\frac{\alpha}{2\pi}\left[\frac{1+(1-y)^2}{y}\ln
\frac{Q_{\rm max}^2}{Q_{\rm min}^2}
 +2m_e^2y\left(\frac{1}{Q_{\rm min}^2}-\frac{1}{Q_{\rm max}^2} \right) \right],
\end{equation}
where $m_e$ is the electron mass, $Q_{\rm min}^2=m_e^2 y^2/(1-y)$ is the
minimum value of the photon virtuality allowed by kinematics, and its maximum
value $Q_{\rm min}^2$ is determined by the experimental conditions, to be
$Q_{\rm min}^2=1$~GeV$^2$ in
Refs.~\cite{H1_data1,H1_data2,ZEUS_data2,ZEUS_data3}.
In the case of resolved photoproduction, $f_{j/\gamma}(x_2,\mu_F)$ is the CPM
PDF of parton $j$ inside the photon.
The case of direct photoproduction is recovered from Eq.~(\ref{sigma_res}) by
setting $f_{j/\gamma}(x_2,\mu_F)=\delta_{\gamma j}\delta(1-x_2)$ and
integrating over $x_2$ using $dy=-y/x_2dx_2$, which follows from the second
equality of Eq.~(\ref{x1x2t1}).

Besides the cross section distributions in $E_T^\gamma$, $\eta^\gamma$,
$E_T^{\rm jet}$, $\eta^{\rm jet}$, $\Delta\phi$, and $y$ given by
Eq.~(\ref{sigma_res}), also other distributions are measured experimentally.
Specifically, the H1 Collaboration also consider the magnitude of the
photon's transverse momentum component orthogonal to the direction of the jet
transverse momentum $p_\perp=E_T^\gamma|\sin(\Delta\phi)|$.
The respective distribution may be obtained from Eq.~(\ref{sigma_res}) via the
replacement $dE_T^\gamma=dp_\perp/|\sin(\Delta\phi)|$.
They also employ the variables
\begin{equation}
 x_p^{\rm LO}=\frac{E_T^\gamma}{2E_p}
\left(e^{\eta^\gamma}+e^{\eta^{\rm jet}}\right),\qquad
 x_\gamma ^{\rm LO}=
 \frac{E_T^\gamma}{2yE_e}\left(e^{-\eta^\gamma}+e^{-\eta^{\rm jet}}\right),
\end{equation}
which, at LO in the CPM, coincide with the fractions of the proton and photon
momentum transferred to the initial-state partons.
The respective distributions follow from Eq.~(\ref{sigma_res}) via the
substitutions $dE_T^\gamma=E_T^\gamma/x_p^{\rm LO}dx_p^{\rm LO}$ and
$dE_T^\gamma=E_T^\gamma/x_\gamma^{\rm LO}dx_\gamma^{\rm LO}$, respectively.
The ZEUS Collaboration uses an alternative variable to probe the
longitudinal-momentum fraction of the parton in the resolved photon, namely
$x_\gamma^{\rm obs}=x_2$, where $x_2$ is given by the second equality in
Eq.~(\ref{x1x2t1}).
The respective distribution emerges from Eq.~(\ref{sigma_res}) via the
replacement $dE_T^\gamma=2yE_ee^{\eta^\gamma}dx_\gamma^{\rm obs}$.
Direct-photoproduction subprocesses at LO in the PRA yield contributions
proportional to $\delta(1-x_\gamma^{\rm obs})$, which are smeared out only by
non-factorisable NLO corrections.

As inputs we use $\alpha=1/137.036$, the LO formula for
$\alpha_s^{(n_f)}(\mu_R)$ with $\Lambda_{\rm LO}^{(n_f)}=220$~MeV for $n_f=4$
active quark flavors \cite{MRST}, the proton unPDF set derived from the LO
proton PDF set by Martin, Sterling, and Thorne \cite{MRST} with $n_f=4$ as
explained in Ref.~\cite{KMR}, and the LO photon PDF set by Gl\"uck, Reya, and
Vogt \cite{GRV} unless otherwise stated.
To estimate the uncertainty related to the photon PDFs, we also use the sets
of Refs.~\cite{DO_PDF, LAC_PDF, WHIT_PDF, SAS_PDF} as implemented in the
PDF library LHAPDF \cite{LHAPDF}.
For our LO CPM predictions, we use the LO proton PDF set \cite{MRST} mentioned
above.
We choose the factorization and renormalization scales to be
$\mu_F=\mu_R=\xi\max(E_T^\gamma, E_T^{\rm jet})$ and vary the parameter $\xi$
in the range $1/2\le\xi\le2$ about its default value $\xi=1$.

We compare our results with five experimental data sets collected by the H1 and
ZEUS collaborations at HERA~II, which we refer to as H1-2005~\cite{H1_data1},
H1-2010~\cite{H1_data2}, ZEUS-2007~I \cite{ZEUS_data2},
ZEUS-2007~II~\cite{ZEUS_data2}, and ZEUS-2013 \cite{ZEUS_data3}.
The respective kinematic conditions are summarized in
Table~\ref{table_kin}.

\begin{turnpage}
\begin{table}
\caption{\label{table_kin}Kinematic conditions of the HERA~II data sets
\cite{H1_data1,H1_data2,ZEUS_data2,ZEUS_data3}.}
\begin{ruledtabular}
\begin{tabular}{ccccc}
H1-2005 \cite{H1_data1} & H1-2010 \cite{H1_data2} &
ZEUS-2007~I \cite{ZEUS_data2} & ZEUS-2007~II \cite{ZEUS_data2} &
ZEUS-2013 \cite{ZEUS_data3} \\
\hline
5.0~GeV${}<E_T^\gamma<10.0$~GeV & 6.0~GeV${}<E_T^\gamma<15.0$~GeV &
5.0~GeV${}<E_T^\gamma<16.0$~GeV & 7.0~GeV${}<E_T^\gamma<16.0$~GeV &
6.0~GeV${}<E_T^\gamma<15.0$~GeV \\
$-1.0<\eta^{\gamma}<0.9$ & $-1.0<\eta^{\gamma}<2.4$ &
$-0.74<\eta^{\gamma}<1.1$ & $-0.74<\eta^{\gamma}<1.1$ &
$-0.7<\eta^{\gamma}<0.9$ \\
$E_T^{\rm jet}>4.5$~GeV & $E_T^{\rm jet}>4.5$~GeV &
6.0~GeV${}<E_T^{\rm jet}<17.0$ GeV & 6.0~GeV${}<E_T^{\rm jet}<17.0$~GeV &
4.0~GeV${}<E_T^{\rm jet}<35.0$~GeV \\
$-1.0<\eta^{\rm jet}<2.3$ & $-1.3<\eta^{\rm jet}<2.3$ &
$-1.6<\eta^{\rm jet}<2.4$ & $-1.6<\eta^{\rm jet}<2.4$ &
$-1.5<\eta^{\rm jet}<1.8$ \\
$0.2<y<0.7$ & $0.1<y<0.7$ & $0.2<y<0.8$ & $0.2<y<0.8$ & $0.2<y<0.7$
\end{tabular}
\end{ruledtabular}
\end{table}
\end{turnpage}

\begin{figure}[H]
\includegraphics[width=0.9\textwidth]{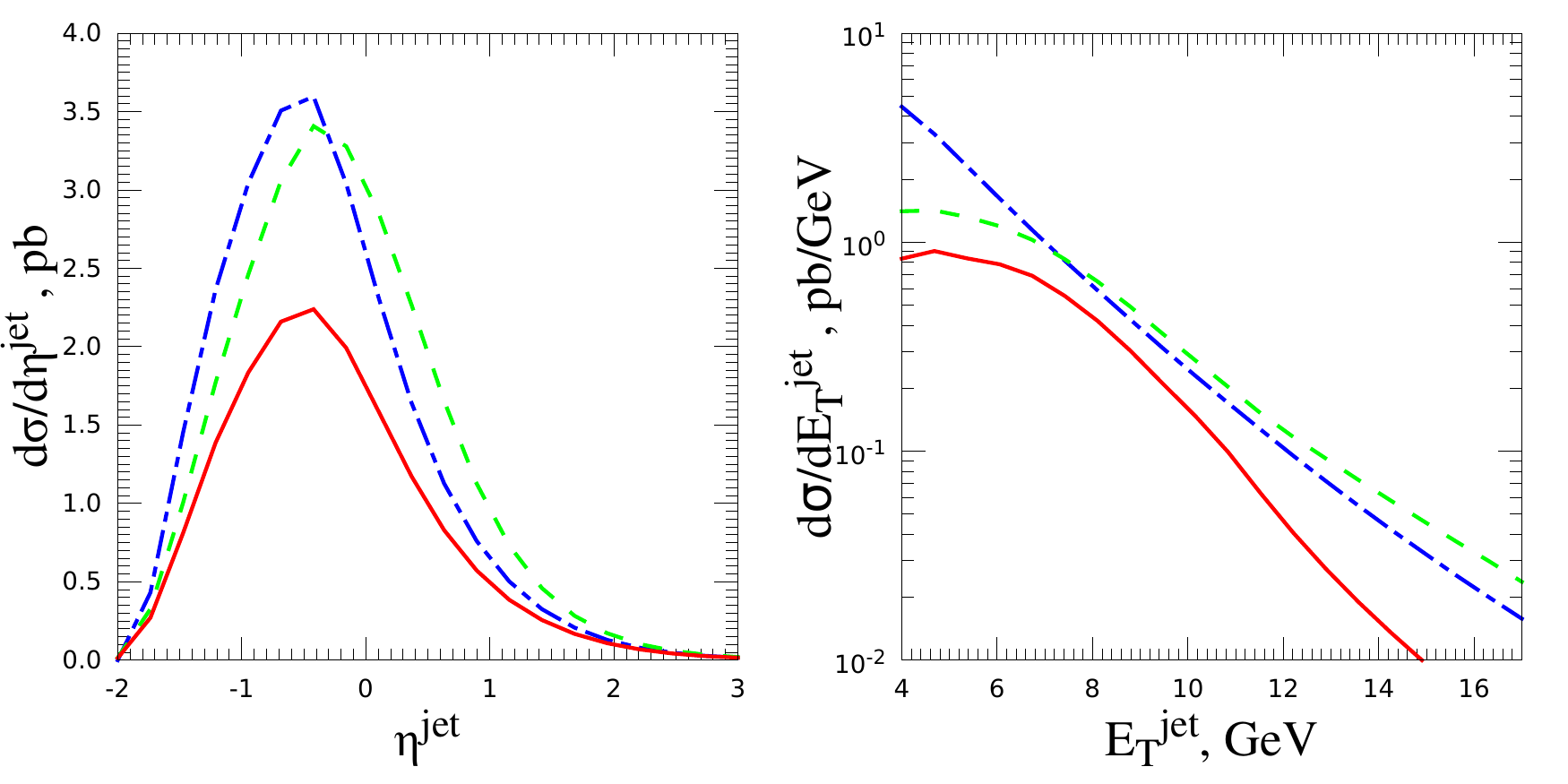}
\caption{\label{fig_comp_boxes}%
(color online).
Contributions due to the loop-induced subprocess in Eq.~(\ref{dir_box}) to the
$\eta^{\rm jet}$ (left panel) and $E_T^{\rm jet}$ (right panel) distributions
of $pe\to\gamma+j+X$ under H1-2005 \cite{H1_data1} kinematic conditions.
The exact PRA results (solid red lines) are compared with the approximate
results obtained by using the CPM box amplitude in the KFA (dashed green lines)
and with the CPM results (dot-dashed blue lines).}
\end{figure}

Prior to comparing with experimental data, we assess the significance of
rigorously evaluating the loop-induced subprocess in Eq.~(\ref{dir_box}) in
the PRA, from Eqs.~(\ref{sigma_res}) and (\ref{box_PRA}), rather than using the
CPM box amplitude in the context of the KFA as was done in Ref.~\cite{MLZ}.
We do this in Fig.~\ref{fig_comp_boxes} for the $\eta^{\rm jet}$ and
$E_T^{\rm jet}$ distributions under H1-2005 \cite{H1_data1} kinematic
conditions.
We observe that, except for small values of $E_T^{\rm jet}$, the approximation
of Ref.~\cite{MLZ} (dashed green lines) is very close the pure CPM result
(dot-dashed blue lines) and significantly overshoots the genuine PRA result
(solid red lines), by as much as 50\% at the peak of the $\eta^{\rm jet}$
distribution.

\begin{figure}[H]
\includegraphics[width=0.9\textwidth]{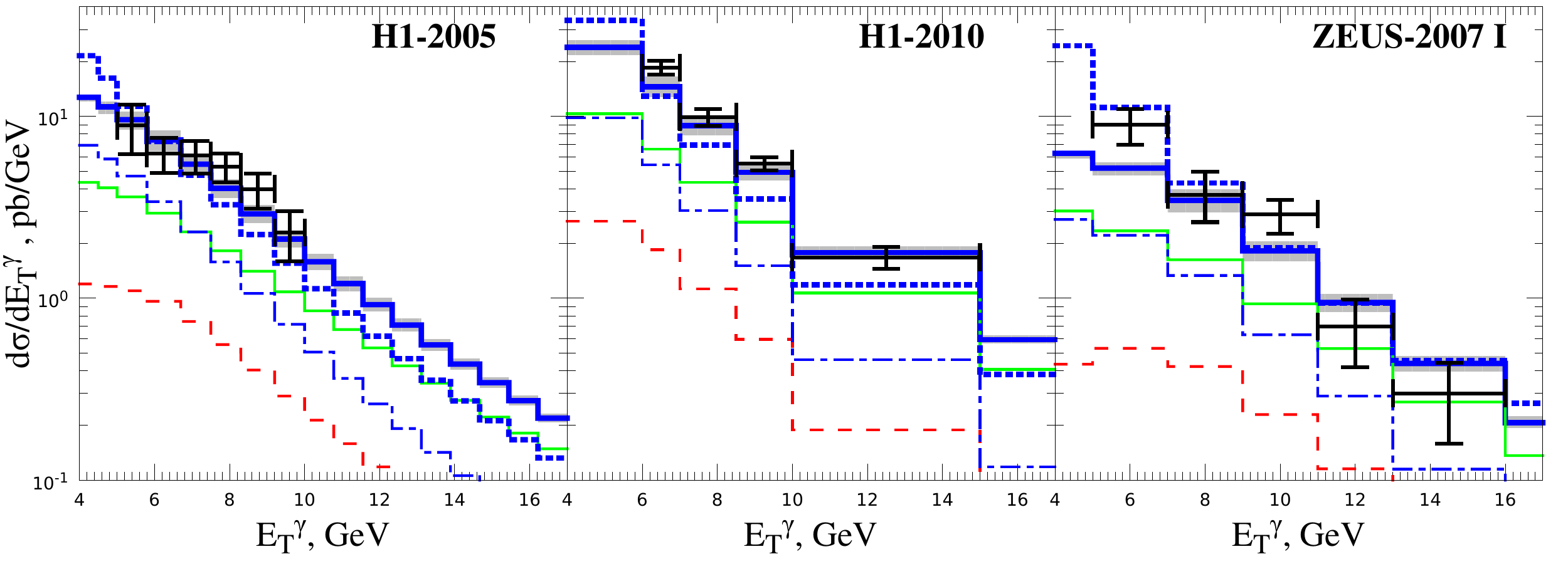}
\caption{\label{fig_p_T_g}%
(color online).
$E_T^\gamma$ distributions of $pe\to\gamma+j+X$ under H1-2005 \cite{H1_data1}
(left panel), H1-2010 \cite{H1_data2} (central panel), and ZEUS-2007~I
\cite{ZEUS_data2} (right panel) kinematic conditions.
The experimental data are compared with LO PRA (boldfaced solid blue lines) and
LO CPM (boldfaced dotted blue lines) predictions.
The theoretical errors in the LO PRA predictions due to the freedom in the
choice of $\xi$ are indicated by the grey bands.
The LO PRA predictions are decomposed into the contributions due to the
partonic subprocesses in Eqs.~(\ref{dir_compton}) (solid green lines),
(\ref{dir_box}) (dashed red lines), and (\ref{res_compton1}) (dot-dashed blue
lines).}
\end{figure}

\begin{figure}[H]
\includegraphics[width=0.9\textwidth]{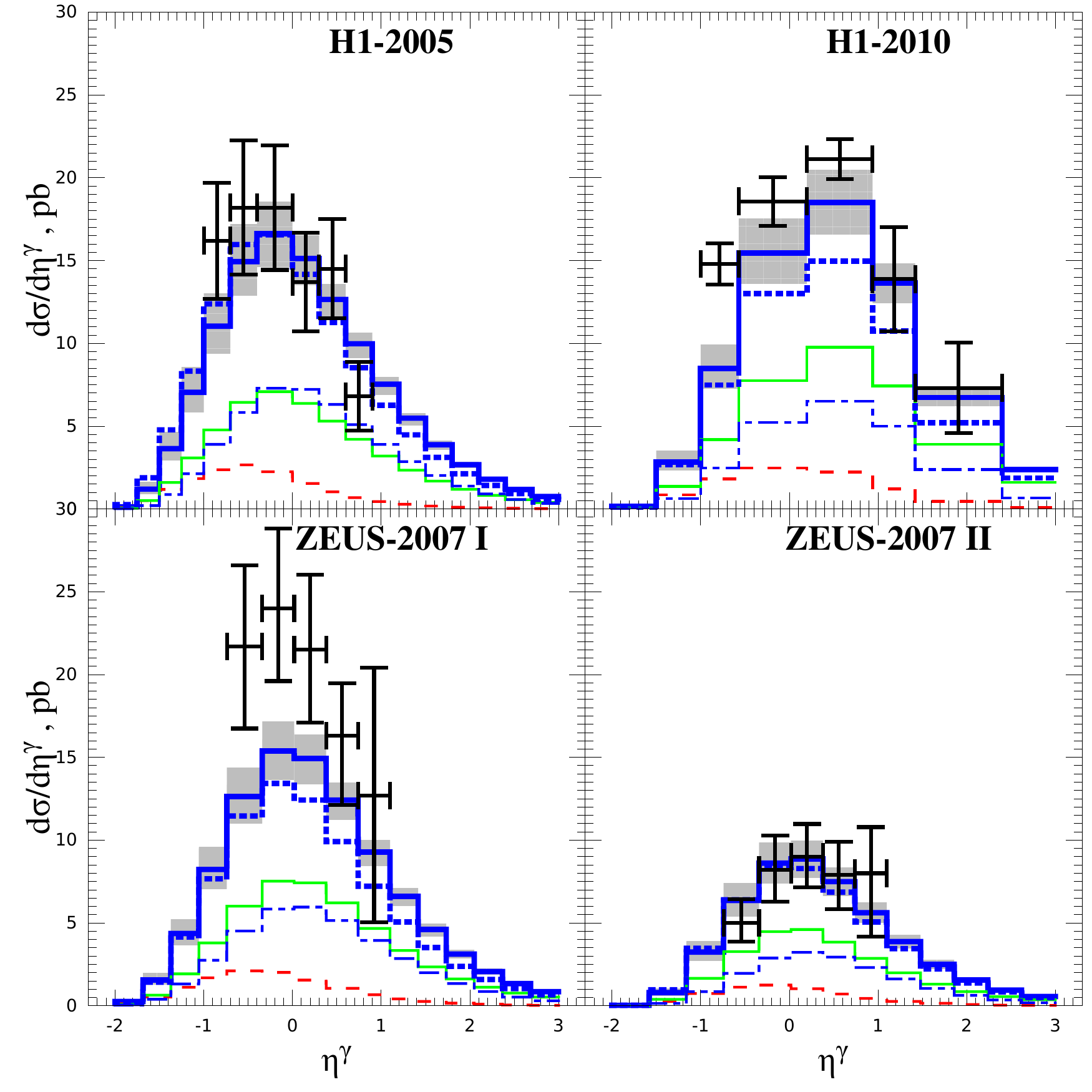}
\caption{\label{fig_eta_g}%
(color online).
$\eta^\gamma$ distributions of $pe\to\gamma+j+X$ under H1-2005 \cite{H1_data1}
(upper left panel), H1-2010 \cite{H1_data2} (upper right panel), ZEUS-2007~I
\cite{ZEUS_data2} (lower left panel), and ZEUS-2007~II \cite{ZEUS_data2} (lower
right panel) kinematic conditions.
Same notation as in Fig.~\ref{fig_p_T_g}.}
\end{figure}

\begin{figure}[H]\
\includegraphics[width=0.9\textwidth]{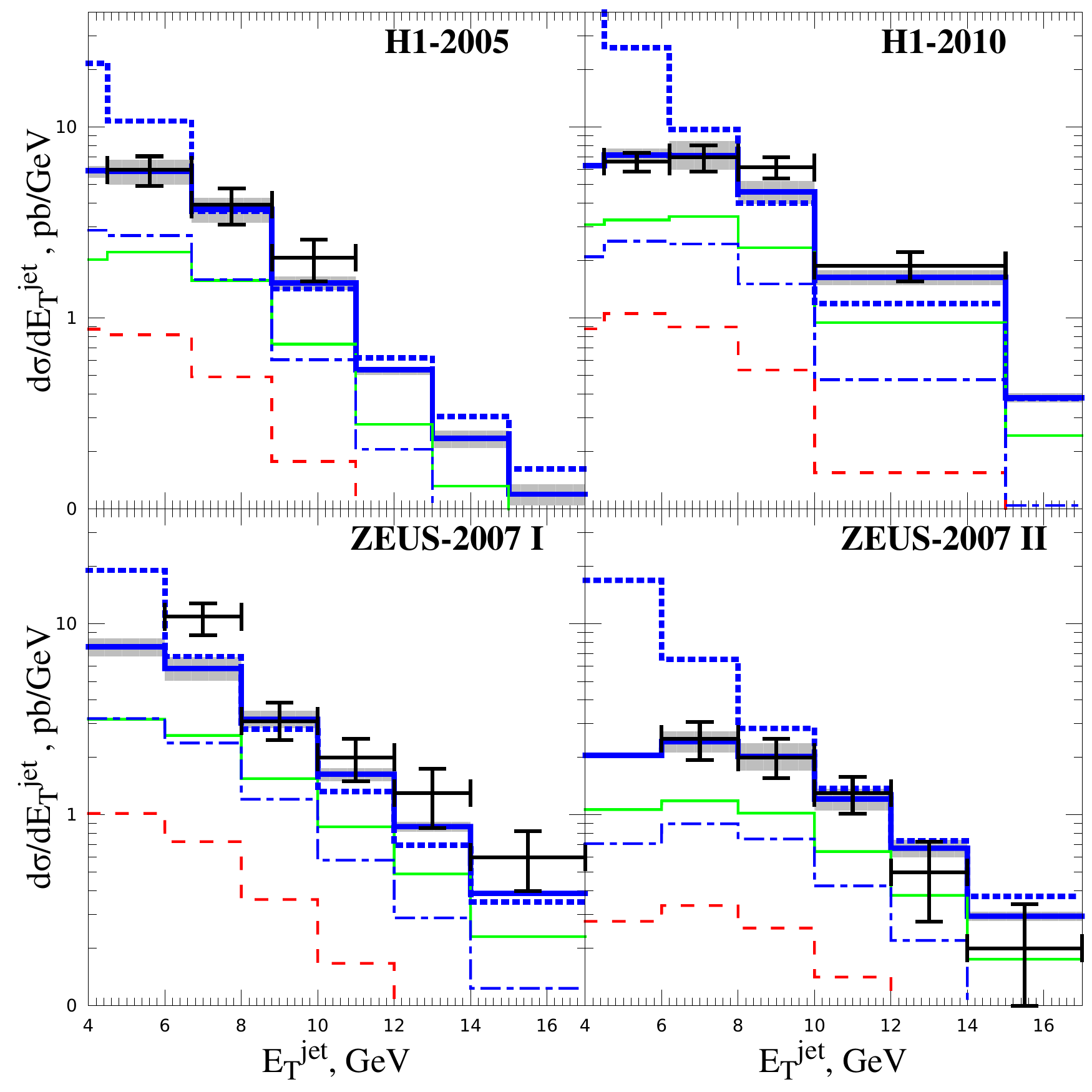}
\caption{\label{fig_p_T_j}%
(color online).
$E_T^{\rm jet}$ distributions of $pe\to\gamma+j+X$ under H1-2005
\cite{H1_data1} (upper left panel), H1-2010 \cite{H1_data2} (upper right
panel), ZEUS-2007~I \cite{ZEUS_data2} (lower left panel), and ZEUS-2007~II
\cite{ZEUS_data2} (lower right panel) kinematic conditions.
Same notation as in Fig.~\ref{fig_p_T_g}.}
\end{figure}

\begin{figure}[H]
\includegraphics[width=0.9\textwidth]{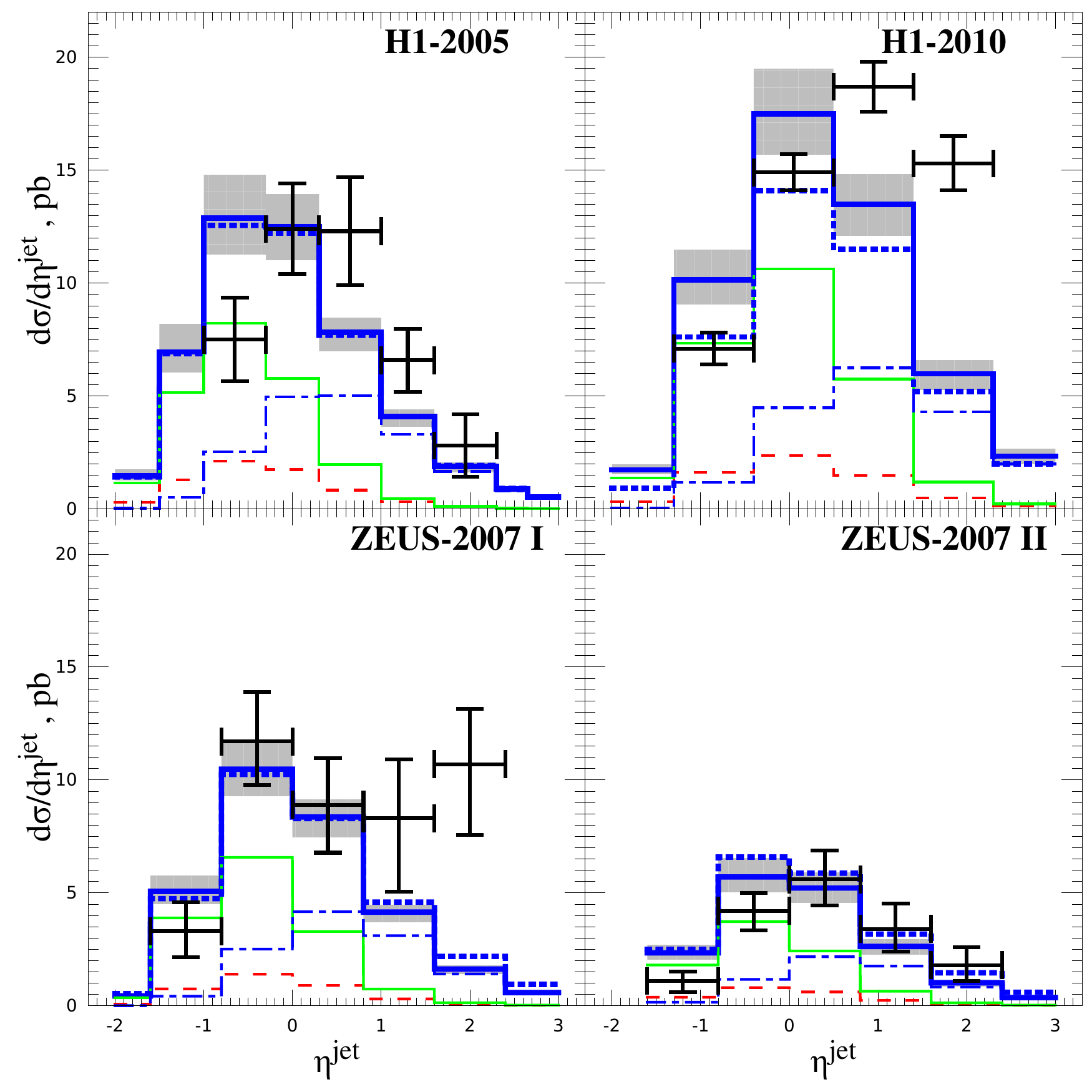}
\caption{\label{fig_eta_j}%
(color online).
$\eta^{\rm jet}$ distributions of $pe\to\gamma+j+X$ under H1-2005
\cite{H1_data1} (upper left panel), H1-2010 \cite{H1_data2} (upper right
panel), ZEUS-2007~I \cite{ZEUS_data2} (lower left panel), and ZEUS-2007~II
\cite{ZEUS_data2} (lower right panel) kinematic conditions.
Same notation as in Fig.~\ref{fig_p_T_g}.}
\end{figure}

\begin{figure}[H]
\includegraphics[width=0.45\textwidth]{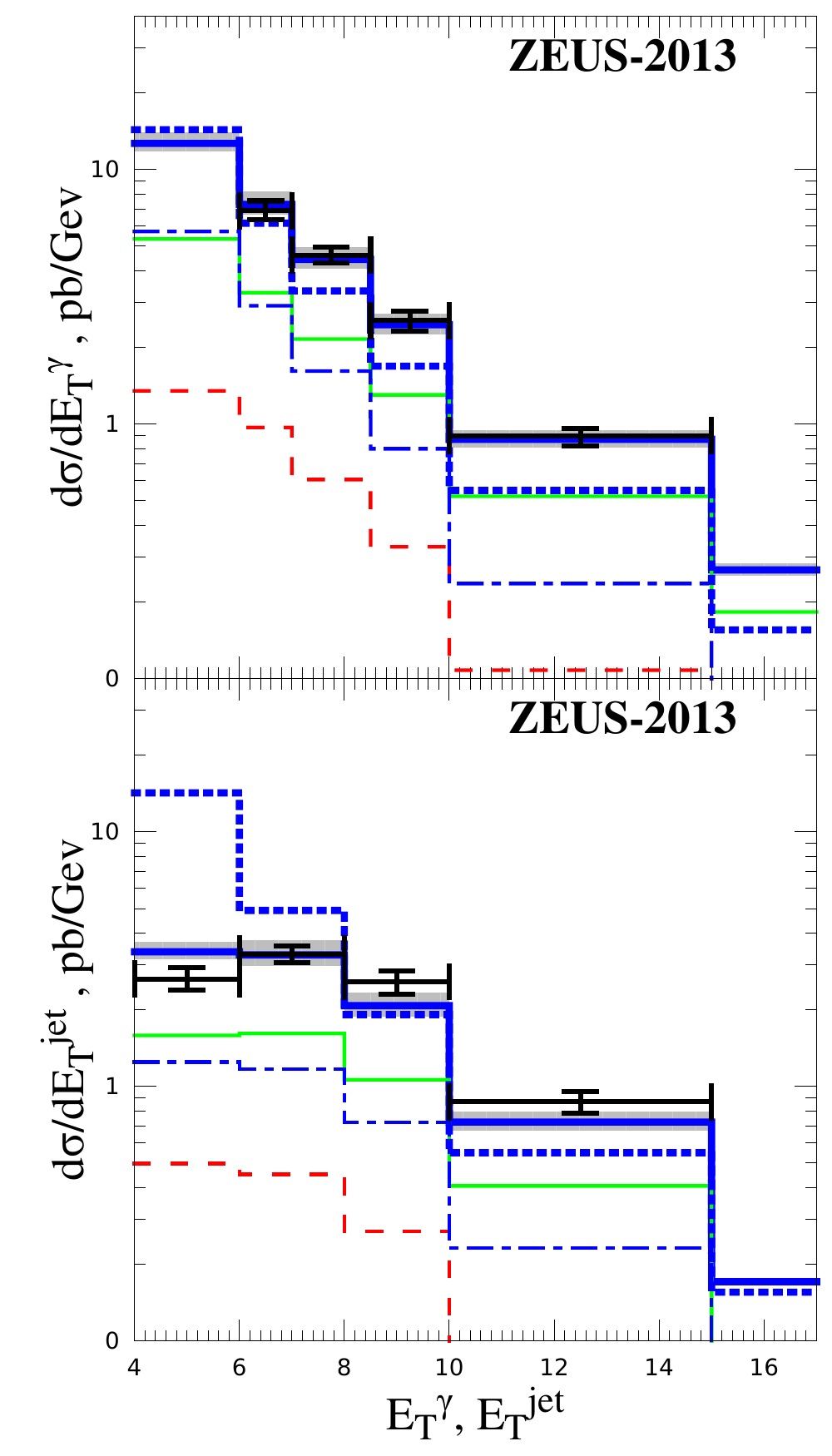}
\includegraphics[width=0.45\textwidth]{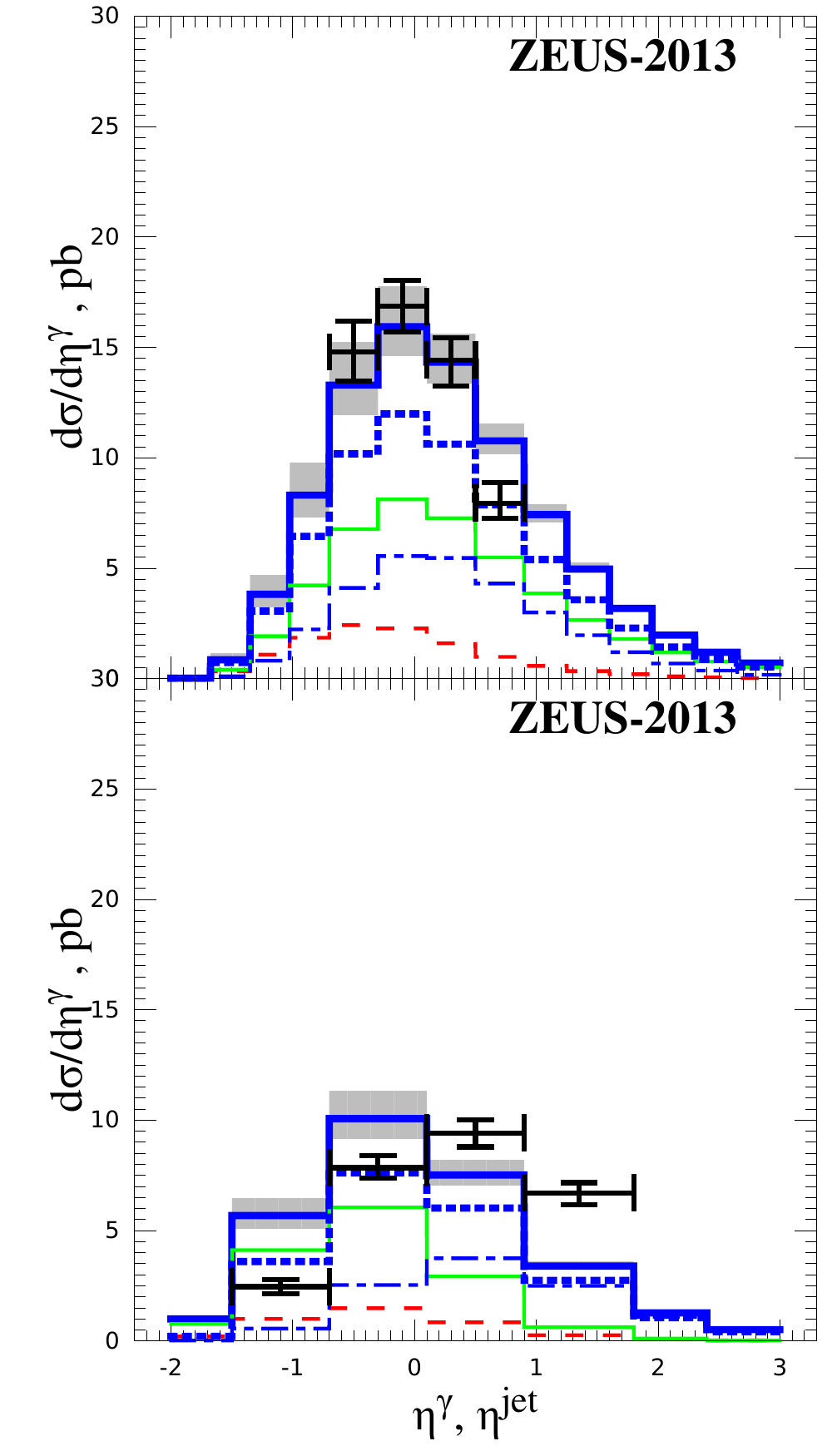}
\caption{\label{fig1_ZEUS-2013}%
(color online).
$E_T^\gamma$ (upper left panel), $\eta^\gamma$ (upper right panel),
$E_T^{\rm jet}$ (lower left panel), and $\eta^{\rm jet}$ (lower right panel)
distributions of $pe\to\gamma+j+X$ under ZEUS-2013 \cite{ZEUS_data3} kinematic
conditions.
Same notation as in Fig.~\ref{fig_p_T_g}.}
\end{figure}

\begin{figure}[H]
\includegraphics[width=0.9\textwidth]{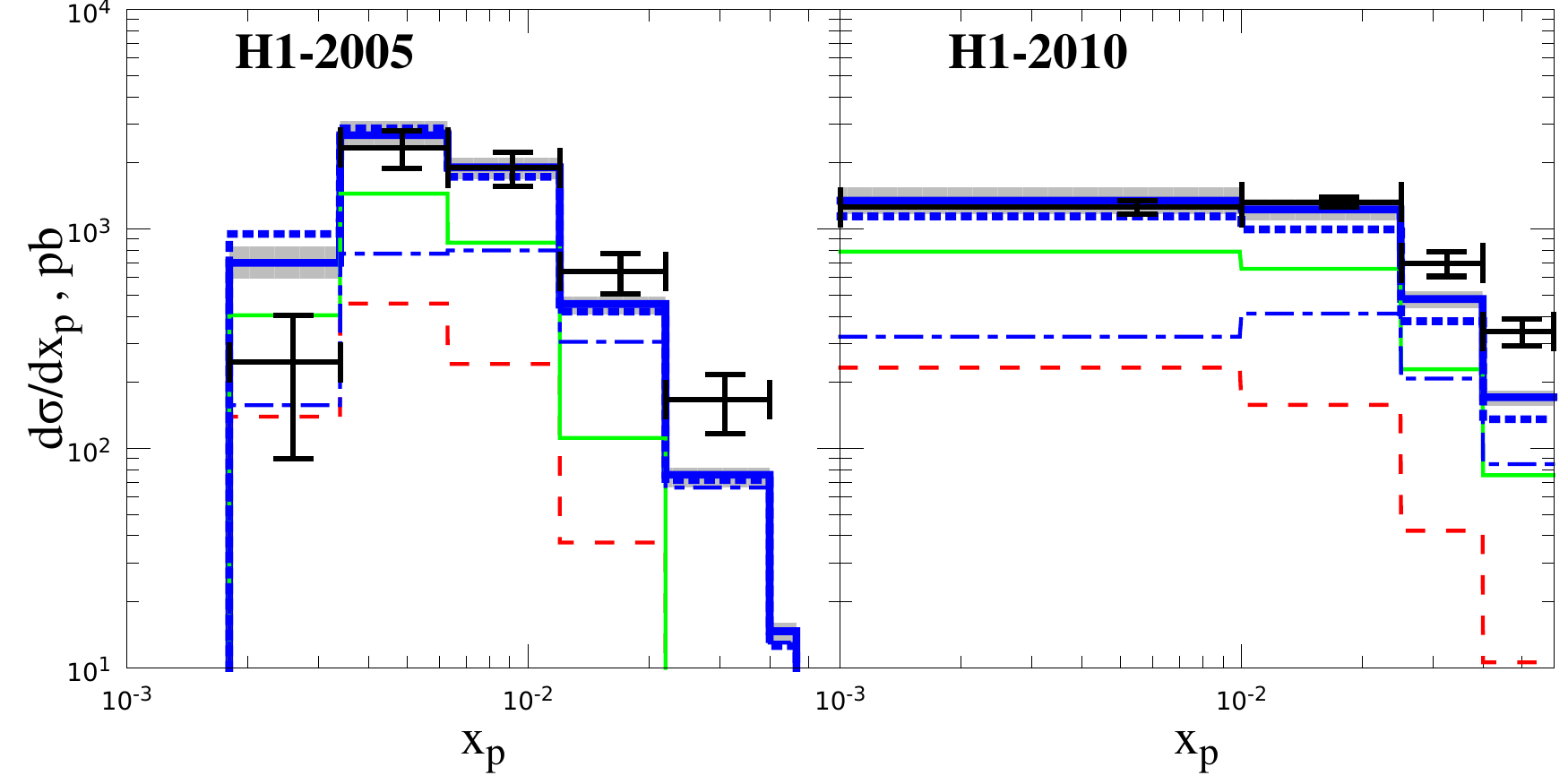}
\caption{\label{fig_x_p}%
(color online).
$x_p^{\rm LO}$ distributions of $pe\to\gamma+j+X$ under H1-2005 \cite{H1_data1}
(left panel) and H1-2010 \cite{H1_data2} (right panel) kinematic conditions.
Same notation as in Fig.~\ref{fig_p_T_g}.}
\end{figure}

\begin{figure}[H]
\includegraphics[width=0.9\textwidth]{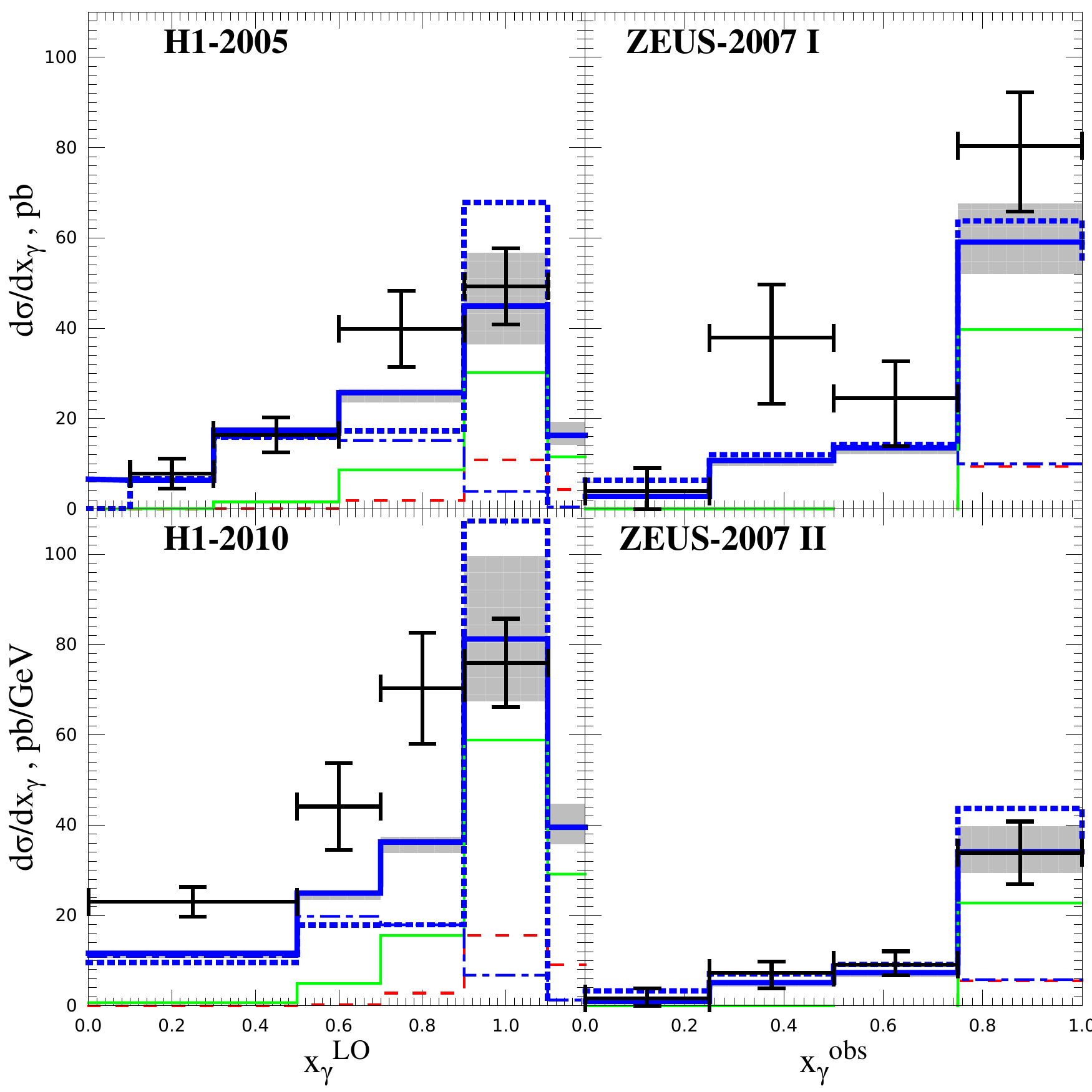}
\caption{\label{fig_x_gam}%
(color online).
$x_\gamma^{\rm LO}$ distributions of $pe\to\gamma+j+X$ under H1-2005
\cite{H1_data1} (upper left panel) and H1-2010 \cite{H1_data2} (lower left
panel) kinematic conditions and $x_\gamma^{\rm obs}$ distributions of
$pe\to\gamma+j+X$ under ZEUS-2007~I \cite{ZEUS_data2} (upper right panel) and
ZEUS-2007~II \cite{ZEUS_data2} (lower right panel) kinematic conditions.
Same notation as in the Fig.~\ref{fig_p_T_g}.}
\end{figure}

\begin{figure}[H]
\includegraphics[width=0.9\textwidth]{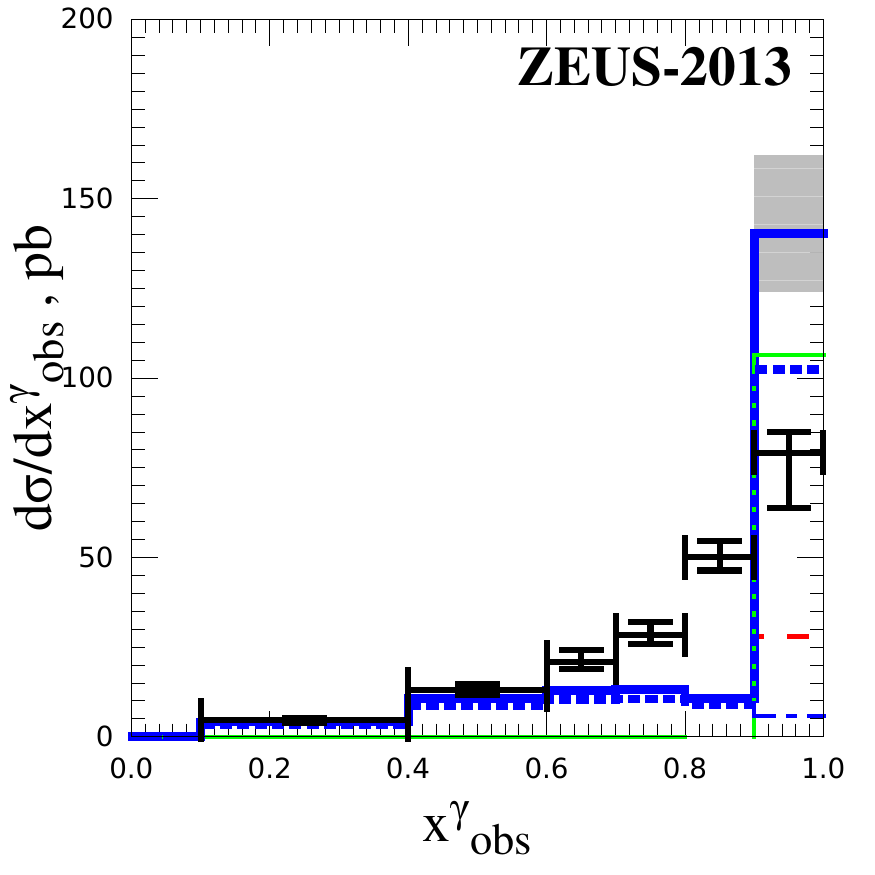}
\caption{\label{fig2_ZEUS-2013}%
(color online).
$x_\gamma^{\rm obs}$ distribution of $pe\to\gamma+j+X$ under ZEUS-2013
\cite{ZEUS_data3} kinematic conditions.
Same notation as in the Fig.~\ref{fig_p_T_g}.}
\end{figure}

\begin{figure}[H]
\includegraphics[width=0.9\textwidth]{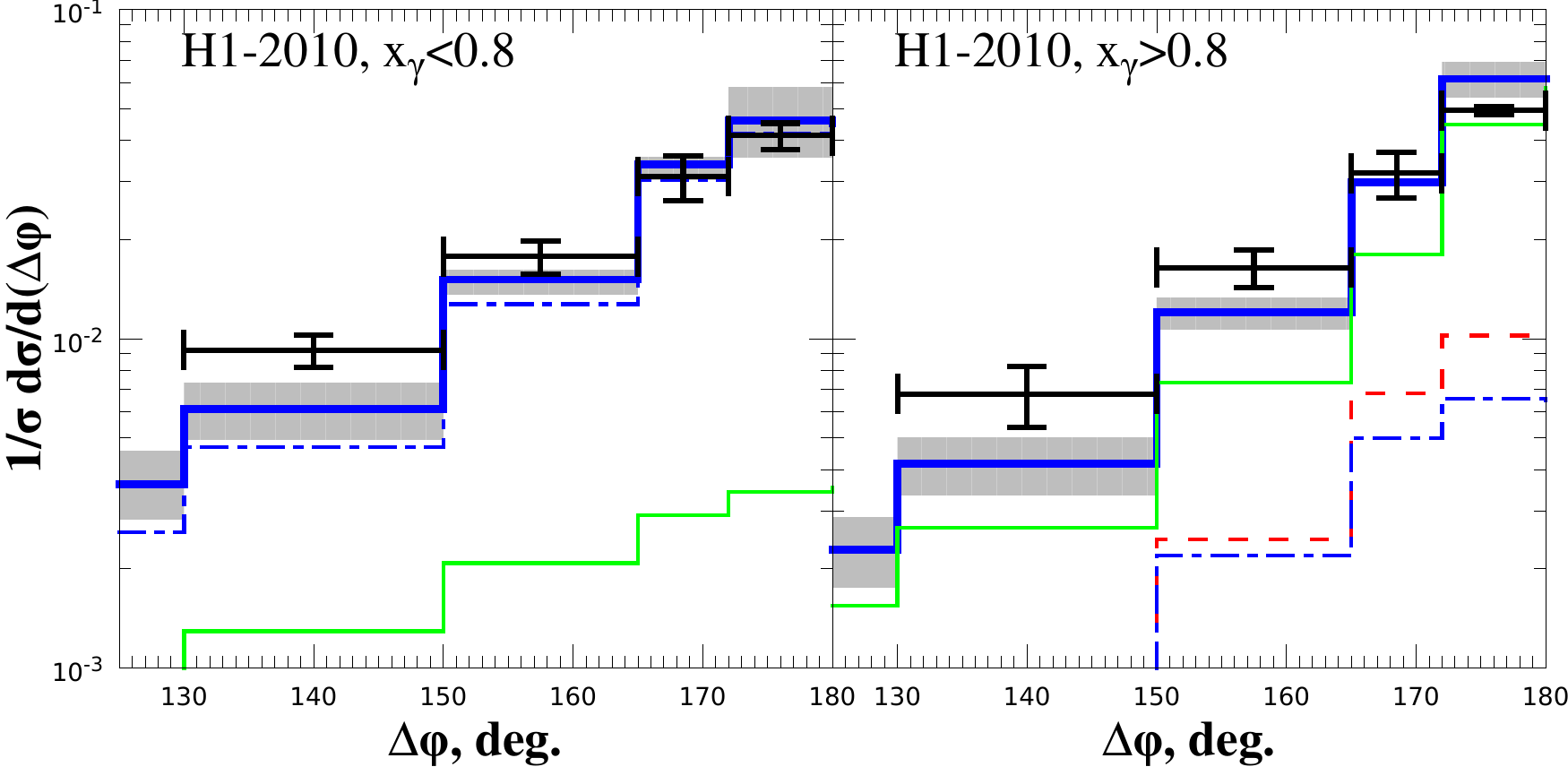}
\caption{\label{fig_phi}%
(color online).
Normalized $\Delta\phi$ distributions of $pe\to\gamma+j+X$ under H1-2010
\cite{H1_data2} kinematic conditions for $x_\gamma^{\rm LO}<0.8$ (left panel)
and $x_\gamma^{\rm LO}>0.8$ (right panel).
Same notation as in the Fig.~\ref{fig_p_T_g}.}
\end{figure}

\begin{figure}[H]
\includegraphics[width=0.9\textwidth]{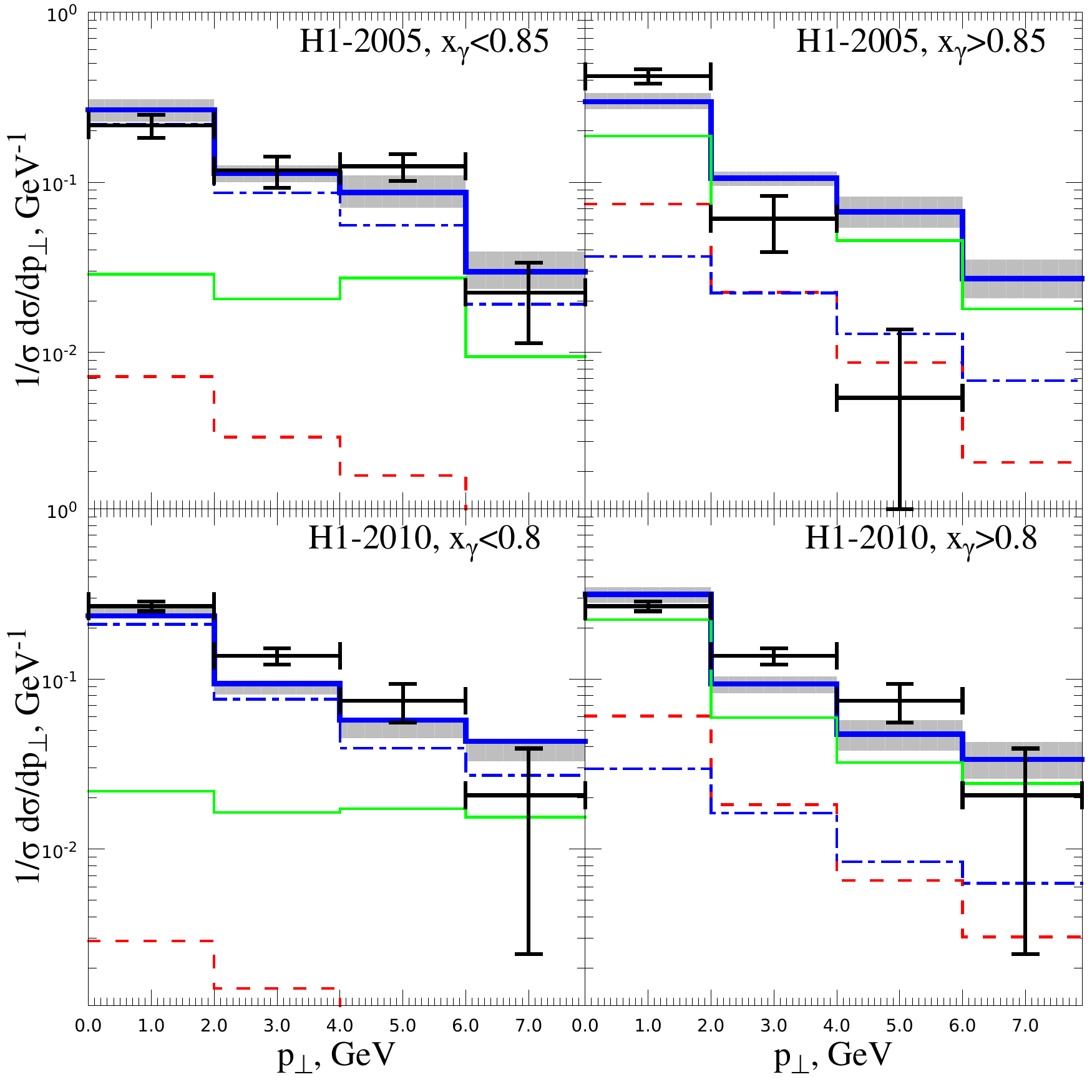}
\caption{\label{fig_p_per}%
(color online).
Normalized $p_\perp$ distributions of $pe\to\gamma+j+X$ under H1-2005
\cite{H1_data1} kinematic conditions for $x_\gamma^{\rm LO}<0.85$ (upper left
panel) and $x_\gamma^{\rm LO}>0.85$ (upper right panel) and under H1-2010
\cite{H1_data2} kinematic conditions for $x_\gamma^{\rm LO}<0.8$ (lower left
panel) and $x_\gamma^{\rm LO}>0.8$ (lower right panel).
Same notation as in the Fig.~\ref{fig_p_T_g}.}
\end{figure}

We now turn to the comparisons with the HERA~II data
\cite{H1_data1,H1_data2,ZEUS_data2,ZEUS_data3}.
Specifically, we consider the $E_T^\gamma$, $\eta^\gamma$, $E_T^{\rm jet}$, and
$\eta^{\rm jet}$  distributions of Refs.~\cite{H1_data1,H1_data2,ZEUS_data2} in
Figs.~\ref{fig_p_T_g}, \ref{fig_eta_g}, \ref{fig_p_T_j}, and \ref{fig_eta_j},
respectively; the same distributions of Ref.~\cite{ZEUS_data3} in
Fig.~\ref{fig1_ZEUS-2013};
the $x_p^{\rm LO}$ distributions of Refs.~\cite{H1_data1,H1_data2} in
Fig.~\ref{fig_x_p};
the $x_\gamma^{\rm LO}$ distributions of Refs.~\cite{H1_data1,H1_data2} and the
$x_\gamma^{\rm obs}$ distributions of Ref.~\cite{ZEUS_data2} in
Fig.~\ref{fig_x_gam};
the $x_\gamma^{\rm obs}$ distribution of Ref.~\cite{ZEUS_data3} in
Fig.~\ref{fig2_ZEUS-2013};
the normalized $\Delta\phi$ distributions of Ref.~\cite{H1_data2} in
Fig.~\ref{fig_phi}; and
the normalized $p_\perp$ distributions of Refs.~\cite{H1_data1,H1_data2} in
Fig.~\ref{fig_p_per}.
In each figure, the LO PRA (boldfaced solid blue lines) predictions are
decomposed into the contributions due to the partonic subprocesses in
Eqs.~(\ref{dir_compton}) (solid green lines), (\ref{dir_box}) (dashed red
lines), and (\ref{res_compton1}) (dot-dashed blue lines) and compared with the
LO CPM predictions (boldfaced dotted blue lines).
The theoretical errors in the LO PRA predictions due to the freedom in the
choice of $\xi$ are indicated by the grey bands.
The normalization factors $\sigma$ in Figs.~\ref{fig_phi} and \ref{fig_p_per}
are evaluated using the corresponding $x_\gamma^{\rm LO}$ cuts.
Comparisons of the experimental data
\cite{H1_data1,H1_data2,ZEUS_data2,ZEUS_data3}
with NLO CPM predictions \cite{NLO_FGH,NLO_KZ} may be found
for the $E_T^\gamma$ distribution
in Fig.~4(c) of Ref.~\cite{H1_data1},
in Fig.~7(a) of Ref.~\cite{H1_data2},
in Fig.~5(a) (ZEUS-2007~I) of Ref.~\cite{ZEUS_data2},
and in Fig.~5(a) of Ref.~\cite{ZEUS_data3};
for the $\eta^\gamma$ distribution
in Fig.~4(d) of Ref.~\cite{H1_data1},
in Fig.~7(b) of Ref.~\cite{H1_data2},
in Figs.~5(b) (ZEUS-2007~I) and 8(a) (ZEUS-2007~II) of Ref.~\cite{ZEUS_data2},
and in Fig.~5(b) of Ref.~\cite{ZEUS_data3};
for the $E_T^{\rm jet}$ distribution
in Fig.~5(a) of Ref.~\cite{H1_data1},
in Fig.~7(c) of Ref.~\cite{H1_data2},
in Figs.~6(a) (ZEUS-2007~I) and 8(b) (ZEUS-2007~II) of Ref.~\cite{ZEUS_data2},
and in Fig.~6(a) of Ref.~\cite{ZEUS_data3};
for the $\eta^{\rm jet}$ distribution
in Fig.~5(b) of Ref.~\cite{H1_data1},
in Fig.~7(d) of Ref.~\cite{H1_data2},
in Figs.~6(b) (ZEUS-2007~I) and 8(c) (ZEUS-2007~II) of Ref.~\cite{ZEUS_data2},
and in Fig.~6(b) of Ref.~\cite{ZEUS_data3};
for the $x_p^{\rm LO}$ distribution
in Fig.~5(d) of Ref.~\cite{H1_data1} and
in Fig.~8(b) of Ref.~\cite{H1_data2};
for the $x_\gamma^{\rm LO}$ or $x_\gamma^{\rm obs}$ distributions
in Fig.~5(c) of Ref.~\cite{H1_data1},
in Fig.~8(a) of Ref.~\cite{H1_data2},
in Figs.~7 (ZEUS-2007~I) and 9 (ZEUS-2007~II) of Ref.~\cite{ZEUS_data2},
and in Fig.~7 of Ref.~\cite{ZEUS_data3};
for the $\Delta\phi$ distribution in Figs.~9(a) ($x_\gamma^{\rm LO}>0.8$) and
9(c) ($x_\gamma^{\rm LO}<0.8$) of Ref.~\cite{H1_data2};
and for the $p_\perp$ distribution in Figs.~6(c) ($x_\gamma^{\rm LO}<0.85$) and
6(d) ($x_\gamma^{\rm LO}>0.85$) of Ref.~\cite{H1_data1} and in
Figs.~9(b) ($x_\gamma^{\rm LO}>0.8$) and 9(d) ($x_\gamma^{\rm LO}<0.8$) in
Ref.~\cite{H1_data2}.

We first assess the relative importance of the LO PRA contributions due to the
partonic subprocesses in Eqs.~(\ref{dir_compton}), (\ref{dir_box}), and
(\ref{res_compton1}).
When $x_\gamma^{\rm LO}$ or $x_\gamma^{\rm obs}$ are not constrained, as in
Figs.~\ref{fig_p_T_g}--\ref{fig_x_p}, then the partonic subprocesses in
Eqs.~(\ref{dir_compton}) and (\ref{res_compton1}) compete with each other,
while the one in Eq.~(\ref{dir_box}) is of minor importance.
Obviously, the loop suppression of the latter is insufficiently compensated by
the dominance of the gluon unPDF over the quark unPDFs.
This feature is more pronounced in the PRA than in the CPM or in the
approximation of Ref.~\cite{MLZ}, as we have seen in
Fig.~\ref{fig_comp_boxes}.
Looking at Figs.~\ref{fig_x_gam} and \ref{fig2_ZEUS-2013}, we observe that
resolved photoproduction, which essentially proceeds via the partonic
subprocess in Eq.~(\ref{res_compton1}), dominates for $x_\gamma^{\rm LO}<0.9$
after H1-2005 \cite{H1_data1} or H1-2010 \cite{H1_data2} cuts, for
$x_\gamma^{\rm obs}<0.75$ after ZEUS-2007~I \cite{ZEUS_data2} and ZEUS-2007~II
\cite{ZEUS_data2} cuts, and for $x_\gamma^{\rm obs}<0.9$ after ZEUS-2013
\cite{ZEUS_data3} cuts.
This is also reflected in Figs.~\ref{fig_phi} and \ref{fig_p_per}, where
the LO PRA predictions for $x_\gamma^{\rm LO}<0.85$ \cite{H1_data1} and
$x_\gamma^{\rm LO}<0.8$ \cite{H1_data2} are almost exhausted by the
contribution due the partonic subprocess in Eq.~(\ref{res_compton1}).
By contrast, the partonic subprocesses of direct photoproduction in
Eqs.~(\ref{dir_compton}) and (\ref{dir_box}) only contribute to the utmost
$x_\gamma^{\rm obs}$ bins in Figs.~\ref{fig_x_gam} and \ref{fig2_ZEUS-2013}.
In order for this peak to be smeared out, one needs to include $2\to3$
subprocesses of direct photoproduction at NLO giving rise to an additional jet
in the central region of rapidity.
From the lower left panel in Fig.~\ref{fig_x_gam}, we observe that the LO PRA
prediction typically undershoots the H1-2010 \cite{H1_data2} data by a factor
of two in the range $x_\gamma^{\rm LO}<0.9$.
The same undershoot would show up in the left panel of Fig.~\ref{fig_phi} and
in the lower left panel of Fig.~\ref{fig_p_per} if it were not for the
normalizations of the $\Delta\phi$ and $p_\perp$ distributions shown there.

Next we compare the LO PRA predictions with the LO CPM ones.
From Figs.~\ref{fig_p_T_g}, \ref{fig_p_T_j}, and \ref{fig1_ZEUS-2013}, we
observe that the $E_T^\gamma$ and $E_T^{\rm jet}$ distributions generally fall
off more steeply in the CPM and significantly overshoot the PRA distributions
at small values of $E_T^\gamma$ and $E_T^{\rm jet}$.
This may be attributed to the fact that the singular behavior of the
partonic cross sections for $E_T^\gamma\to0$ or $E_T^{\rm jet}\to0$ in the CPM
is washed out by the PRA dynamics and the $k_T$ smearing via the unPDFs.
From Fig.~\ref{fig_p_T_g}, we also learn that the LO CPM predictions for the
H1-2005 \cite{H1_data1} or H1-2010 \cite{H1_data2} experimental conditions
undershoot the LO PRA ones for $E_T^\gamma>6$~GeV.
Consequently, the same is true for the H1-2010 \cite{H1_data2}
$\eta^\gamma$ and $\eta^{\rm jet}$ distributions in Figs.~\ref{fig_eta_g} and
\ref{fig_eta_j}, respectively, because of the very cut $E_T^\gamma>6$~GeV.
A similar observation can be made in Fig.~\ref{fig1_ZEUS-2013} for the
ZEUS-2013 \cite{ZEUS_data3} situation:
The LO CPM $E_T^\gamma$ distribution undershoots the LO PRA one for
$E_T^\gamma>6$~GeV, which carries over the $\eta^\gamma$ and $\eta^{\rm jet}$
distributions being subject to this very cut.
Since the prompt photon and the jet are strictly back to back at LO in the CPM,
the respective contributions to the $\Delta\phi$ and $p_\perp$ distributions
are zero, as may be seen from Figs.~\ref{fig_phi} and \ref{fig_p_per}.

At this point, we estimate the theoretical uncertainty due to the imperfect
knowledge of the photon PDFs.
We do this by recalculating the $x_\gamma^{\rm LO}$ distributions in
Fig.~\ref{fig_x_gam}, which are particularly sensitive probes of this, using
four alternative photon PDF sets \cite{DO_PDF,LAC_PDF,WHIT_PDF,SAS_PDF} as
implemented in the PDF library LHAPDF \cite{LHAPDF}.
We find the variation to be $\pm(10$--$20)\%$ in the interval
$0.2<x_\gamma^{\rm LO}<0.9$ and below $\pm10\%$ in the utmost bin.

Finally, we compare the predictions at LO in the PRA and at NLO in the CPM
\cite{NLO_FGH,NLO_KZ} with respect to their abilities to describe the
experimental data \cite{H1_data1,H1_data2,ZEUS_data2,ZEUS_data3}.
We find their overall performances to be comparable, except that, at LO in
the PRA, the peak positions of the $\eta^{\rm jet}$ distributions are generally
too small and the $x_p^{\rm LO}$ distributions tend to be too small in the
utmost bins.
On the other hand, the CPM at NLO significantly undershoots the measured
$\Delta\phi$ distribution for $x_\gamma^{\rm LO}<0.8$ in the utmost bin, where
the PRA at LO does an excellent job.
However, these comparisons have to be taken with a grain of salt because the
NLO CPM predictions presented in
Refs.~\cite{H1_data1,H1_data2,ZEUS_data2,ZEUS_data3} include corrections due to
hadronization and multiple interactions, which are beyond the scope of our
present analysis.

\section{Conclusions}
\label{sec:concl}

We studied prompt-photon plus jet associated photoproduction at LO in the PRA,
treating the quarks and gluons inside the proton as Reggeized particles and
allowing for the incoming photon to be resolved.
We also included the loop-induced subprocess in Eq.~(\ref{dir_box}), which
was treated in the PRA accounting for the off-shellness of the Reggeon in a
manifestly gauge-invariant way for the first time.
We performed detailed comparisons with experimental data taken by the H1
\cite{H1_data1,H1_data2} and ZEUS \cite{ZEUS_data2,ZEUS_data3} collaborations
at HERA~II, which come as cross section distributions in $E_T^\gamma$,
$\eta^\gamma$, $E_T^{\rm jet}$, $\eta^{\rm jet}$, $x_p^{\rm LO}$,
$x_\gamma^{\rm LO}$, $x_\gamma^{\rm obs}$, $\Delta\phi$, and $p_\perp$.
We generally found good agreement, which indicates that factorizable
higher-order corrections are significant here.

\section*{Acknowledgments}

We thank A.~Iudin and K.~Nowak for a clarifying communication concerning
Ref.~\cite{ZEUS_data2},
E.~Lohrmann for drawing our attention to Ref.~\cite{ZEUS_data3},
and A. Kotikov and O. Veretin for useful comments on the box contribution.
The work of M.A.N. was supported in part by the German Academic Exchange
Service DAAD and the Ministry of Science and Education of the Russian
Federation through Michail Lomonosov Grant No.~A/12/75163 and by the
Dynasty Foundation through a Grant from the Graduate Students Stipend Program.
The work of M.A.N. and V.A.S. was supported in part by the Russian Foundation
for Basic Research through Grant No.\ 14-02-00021.
This work was supported in part by the German Federal Ministry for Education
and Research BMBF through Grant No.\ 05H12GUE.

\begin{appendix}

\section{Box amplitude}
\label{sec:appA}

In this Appendix, we present the independent helicity amplitudes in
Eq.~(\ref{Rgamma}) of the partonic subprocess in Eq.~(\ref{dir_box}).
They, may be written as
\begin{eqnarray}
{\cal M}(R+,++)&=&{\cal M}\left(t,u,t_1,\lbrace f^{(1)}_i\rbrace, {\cal R}_1 \right),
\label{HA1}\\
{\cal M}(R+,+-)&=&{\cal M}\left(s,t,t_1,\lbrace f^{(2)}_i\rbrace, {\cal R}_2\right),
\label{HA2}\\
{\cal M}(R+,-+)&=&{\cal M}\left(s,u,t_1,\lbrace f^{(3)}_i\rbrace, {\cal R}_3\right),
\label{HA3}\\
{\cal M}(R+,--)&=&  \frac{i\pi^2 4\sqrt{2}}{u\Delta} (t+u) \gamma_1,
\label{HA4}
\end{eqnarray}
where
\begin{eqnarray}
{\cal M}\left(t,u,t_1,\left\lbrace f_i \right\rbrace, {\cal R} \right) &=&\frac{i\pi^2}{\sqrt{2}
\Delta^3(t+u)}\left\{ f_1 \left[B_0(t)-B_0(-t_1) \right]+ f_2 \left[B_0(u)-B_0(-t_1) \right] \right.\nonumber \\
 &&{}+ \left. f_3 E(t_1,t,u)+{\cal R} \right\},
\end{eqnarray}
with
\begin{equation}
 E(t_1,t,u)=t C_0(t)+ u C_0(u)+ (t+t_1) C_0(-t_1,t)+ (u+t_1) C_0(-t_1,u)-tu D_0(-t_1,t,u).
\end{equation}
In the notation of Ref.~\cite{K_Ellis}, the scalar one-loop integrals are
defined as
  \begin{eqnarray}
  B_0(p^2_1)&=&I_2^D(p_1^2;0,0),\nonumber\\
  C_0(p_3^2)&=&I_3^D(0,0,p_3^2;0,0,0),\nonumber\\
  C_0(p_2^2,p_3^2)&=&I_3^D(0,p_2^2,p_3^2;0,0,0),\nonumber\\
  D_0(s_{12},s_{23})&=&I_4^D(0,0,0,0;s_{12},s_{23};0,0,0,0),\nonumber\\
  D_0(p_4^2,s_{12},s_{23})&=&I_4^D(0,0,0,p_4^2;s_{12},s_{23};0,0,0,0).
  \end{eqnarray}
The coefficients pertaining to Eq.~(\ref{HA1}) read:
\begin{eqnarray}
f_1^{(1)}&=&\frac{-i t^2}{2(t+t_1)^2} \left\{2 (s+2 u)   \left(t+t_1\right)  (t+u)^2\gamma _1 +
   4i s u^2  \left[2 t \left(t+t_1\right)-u
   t_1\right]\sqrt{t_1} \right.\nonumber \\
   &&{}+ \left. u\left[s^2 \left(s+t_1\right) +3 su \left(s-t_1\right) +2
   u^2 \left(s-t_1\right)\right] i\gamma _2\right\},\nonumber\\
f_2^{(1)} &=& \frac{-i t u}{2(u+t_1)^2} \left\{2 (s+2 t)\left(u+t_1\right)(t+u)^2\gamma _1+4i s t u \left[t t_1-2 u
   \left(u+t_1\right)\right]\sqrt{t_1} \right.\nonumber\\
   &&{}+ \left. u\left[s^3+s^2\left(3 t+t_1\right) +st
   \left(2 t-3 t_1\right) -2 t^2 t_1\right]i \gamma _2\right\},\nonumber\\
f_3^{(1)} &=& \frac{-it}{4s} \left\{2(t+u)^2\left[t^2+t_1 t+u \left(u+t_1\right)\right] \gamma_1 +4i s t u^2
   (u-t)\sqrt{t_1} \right.\nonumber \\
 &&{}+ \left. u\left[t^3+t^2 \left(u+t_1\right) +tu \left(u-2 t_1\right)
   +u^2 \left(u+t_1\right)\right]i\gamma _2\right\},\nonumber\\
{\cal R}_1&=&\frac{st^2u^2}{(t+t_1)(u+t_1)}\left[(t_1-s)\gamma_2+2s(t-u)\sqrt{t_1}\right],
\end{eqnarray}
where $\gamma_1$ and $\gamma_2$ are defined in Eqs.~(\ref{g1def}) and
(\ref{g2def}), respectively.
The coefficients pertaining to Eq.~(\ref{HA2}) read:
\begin{eqnarray}
f_1^{(2)}&=&\frac{-i s^2 t}{2u} \left[2 (t+u) (2 t+u)  \gamma _1-4itu^2\sqrt{t_1}-u(2 t+u)i\gamma _2\right],\nonumber\\
f_2^{(2)}&=&\frac{i s t^2}{2u(t+t_1)^2} \left\{2 (2 s+u)  \left(t+t_1\right)(t+u)^2 \gamma _1-4i s u^2 \left[u t_1+t
   \left(t+t_1\right)\right]\sqrt{t_1}\right.\nonumber\\
 &&{}-\left. u\left[2 \left(s+t_1\right) s^2+3 su \left(s+t_1\right)+u^2
   \left(s-t_1\right)\right]i \gamma _2\right\},\nonumber\\
f_3^{(2)}&=&\frac{ist}{4u^2} \left\{2 \left[s^2+t_1 s+t \left(t+t_1\right)\right](t+u)^2\gamma_1 +4i s t^2 u^2 \sqrt{t_1}\right.\nonumber \\
 &&{}- \left. u
   \left[u^3+u^2\left(3 t+t_1\right) +tu \left(4 t+t_1\right) +2 t^2 \left(t+t_1\right)\right]i \gamma_2\right\},\nonumber\\
{\cal R}_2&=&-\frac{s^2 t^2 u}{t+t_1}\left(2u\sqrt{t_1}+\gamma _2\right).
\label{HA2a}
\end{eqnarray}
The coefficients pertaining to Eq.~(\ref{HA3}) emerge from Eq.~(\ref{HA2a})
via the substitutions
 \begin{equation}
 t\leftrightarrow u,\qquad
 \sqrt{t_1}\to -\sqrt{t_1},\qquad
 \gamma_1\to \gamma_1\frac{t}{u},
 \end{equation}
 which amounts to permutating the final-state partons.

The modulus square of the hard-scattering amplitude of the partonic subprocess
in Eq.~(\ref{dir_box}) averaged over the spins and colors in the initial state
and summed over those in the final state is then obtained from the helicity
amplitudes in Eqs.~(\ref{HA1})--(\ref{HA4}) as
 \begin{equation}
 \overline{\left\vert{\cal M}(R+\gamma\to g+\gamma) \right\vert^2}
=\frac{\alpha^2\alpha_s^2}{4\pi^4} \left(\sum\limits_q e_q^2\right)^2
\sum_{\lambda_3,\lambda_4=\pm1}
\left\vert{\cal M}(R+,\lambda_3\lambda_4)\right\vert^2.
\label{box_PRA}
\end{equation}
For completeness, we also present the corresponding CPM result
\cite{BoxCPM},
 \begin{eqnarray}
 \overline{\left\vert {\cal M}(g\gamma\to g\gamma) \right\vert ^2} &=& 8\alpha^2 \alpha_s^2 \left(\sum\limits_q e_q^2\right)^2
 \left\lbrace\left\vert {\cal M}(++,++) \right\vert ^2 + \left\vert {\cal M}(-+,-+) \right\vert ^2 + \left\vert {\cal M}(-+,+-)
  \right\vert ^2  \right. \nonumber \\
 &&{}+ \left. \left\vert {\cal M}(++,--) \right\vert ^2 + 4 \left\vert {\cal M}(++,+-) \right\vert ^2 \right\rbrace, \label{box_CPM}
 \end{eqnarray}
where
 \begin{eqnarray}
 {\cal M}(++,--)&=& {\cal M}(++,+-)=-1,\nonumber \\
 {\cal M}(++,++)&=& 1+(2x-1)L_2+\frac{1}{2}\left[x^2+(1-x)^2\right](L_2+\pi^2),
\nonumber \\
 {\cal M}(-+,-+)&=& 1+\left(1-\frac{2}{x}\right)(L_1-\pi i)+\frac{1}{2x^2}\left[1+ (1-x)^2\right]L_1(L_1-2\pi i),\nonumber\\
 {\cal M}(-+,+-)&=&\left.{\cal M}(-+,-+)\right|_{x\to 1-x},
 \end{eqnarray}
 with $L_1=\log[1/(1-x)]$, $L_2=\log[(1-x)/x]$, and $x=-t/s+i0$.

\end{appendix}

\end{document}